\documentclass[twocolumn,letterpaper,aps,prd,superscriptaddress,longbibliography,floatfix]{revtex4-1}

\usepackage{graphicx}	
\usepackage{amsmath}
\usepackage[utf8]{inputenc}
\usepackage{xspace}	

\begin{document}



\title{Single-spin asymmetry of $J/\psi$ production in $p$$+$$p$, 
$p$$+$Al, and $p$$+$Au collisions with transversely 
polarized proton beams at $\sqrt{s_{_{NN}}}=200$ GeV}

\newcommand{\abilene}{Abilene Christian University, Abilene, Texas 79699, USA}
\newcommand{\augie}{Department of Physics, Augustana University, Sioux Falls, South Dakota 57197, USA}
\newcommand{\banaras}{Department of Physics, Banaras Hindu University, Varanasi 221005, India}
\newcommand{\barc}{Bhabha Atomic Research Centre, Bombay 400 085, India}
\newcommand{\baruch}{Baruch College, City University of New York, New York, New York, 10010 USA}
\newcommand{\bnlcoll}{Collider-Accelerator Department, Brookhaven National Laboratory, Upton, New York 11973-5000, USA}
\newcommand{\bnlphys}{Physics Department, Brookhaven National Laboratory, Upton, New York 11973-5000, USA}
\newcommand{\caucr}{University of California-Riverside, Riverside, California 92521, USA}
\newcommand{\charlesczech}{Charles University, Ovocn\'{y} trh 5, Praha 1, 116 36, Prague, Czech Republic}
\newcommand{\chonbuk}{Chonbuk National University, Jeonju, 561-756, Korea}
\newcommand{\cns}{Center for Nuclear Study, Graduate School of Science, University of Tokyo, 7-3-1 Hongo, Bunkyo, Tokyo 113-0033, Japan}
\newcommand{\colorado}{University of Colorado, Boulder, Colorado 80309, USA}
\newcommand{\columbia}{Columbia University, New York, New York 10027 and Nevis Laboratories, Irvington, New York 10533, USA}
\newcommand{\czechtech}{Czech Technical University, Zikova 4, 166 36 Prague 6, Czech Republic}
\newcommand{\debrecen}{Debrecen University, H-4010 Debrecen, Egyetem t{\'e}r 1, Hungary}
\newcommand{\elte}{ELTE, E{\"o}tv{\"o}s Lor{\'a}nd University, H-1117 Budapest, P{\'a}zm{\'a}ny P.~s.~1/A, Hungary}
\newcommand{\eszterhazy}{Eszterh\'azy K\'aroly University, K\'aroly R\'obert Campus, H-3200 Gy\"ongy\"os, M\'atrai \'ut 36, Hungary}
\newcommand{\ewha}{Ewha Womans University, Seoul 120-750, Korea}
\newcommand{\fsu}{Florida State University, Tallahassee, Florida 32306, USA}
\newcommand{\gsu}{Georgia State University, Atlanta, Georgia 30303, USA}
\newcommand{\hiroshima}{Hiroshima University, Kagamiyama, Higashi-Hiroshima 739-8526, Japan}
\newcommand{\howard}{Department of Physics and Astronomy, Howard University, Washington, DC 20059, USA}
\newcommand{\ihepprot}{IHEP Protvino, State Research Center of Russian Federation, Institute for High Energy Physics, Protvino, 142281, Russia}
\newcommand{\illuiuc}{University of Illinois at Urbana-Champaign, Urbana, Illinois 61801, USA}
\newcommand{\inrras}{Institute for Nuclear Research of the Russian Academy of Sciences, prospekt 60-letiya Oktyabrya 7a, Moscow 117312, Russia}
\newcommand{\instpasczech}{Institute of Physics, Academy of Sciences of the Czech Republic, Na Slovance 2, 182 21 Prague 8, Czech Republic}
\newcommand{\isu}{Iowa State University, Ames, Iowa 50011, USA}
\newcommand{\jaea}{Advanced Science Research Center, Japan Atomic Energy Agency, 2-4 Shirakata Shirane, Tokai-mura, Naka-gun, Ibaraki-ken 319-1195, Japan}
\newcommand{\kek}{KEK, High Energy Accelerator Research Organization, Tsukuba, Ibaraki 305-0801, Japan}
\newcommand{\korea}{Korea University, Seoul, 136-701, Korea}
\newcommand{\kurchatov}{National Research Center ``Kurchatov Institute", Moscow, 123098 Russia}
\newcommand{\kyoto}{Kyoto University, Kyoto 606-8502, Japan}
\newcommand{\lawllnl}{Lawrence Livermore National Laboratory, Livermore, California 94550, USA}
\newcommand{\losalamos}{Los Alamos National Laboratory, Los Alamos, New Mexico 87545, USA}
\newcommand{\lund}{Department of Physics, Lund University, Box 118, SE-221 00 Lund, Sweden}
\newcommand{\lyon}{IPNL, CNRS/IN2P3, Univ Lyon, Université Lyon 1, F-69622, Villeurbanne, France}
\newcommand{\maryland}{University of Maryland, College Park, Maryland 20742, USA}
\newcommand{\mass}{Department of Physics, University of Massachusetts, Amherst, Massachusetts 01003-9337, USA}
\newcommand{\michigan}{Department of Physics, University of Michigan, Ann Arbor, Michigan 48109-1040, USA}
\newcommand{\muhlenberg}{Muhlenberg College, Allentown, Pennsylvania 18104-5586, USA}
\newcommand{\nara}{Nara Women's University, Kita-uoya Nishi-machi Nara 630-8506, Japan}
\newcommand{\natmephi}{National Research Nuclear University, MEPhI, Moscow Engineering Physics Institute, Moscow, 115409, Russia}
\newcommand{\newmex}{University of New Mexico, Albuquerque, New Mexico 87131, USA}
\newcommand{\nmsu}{New Mexico State University, Las Cruces, New Mexico 88003, USA}
\newcommand{\ohio}{Department of Physics and Astronomy, Ohio University, Athens, Ohio 45701, USA}
\newcommand{\ornl}{Oak Ridge National Laboratory, Oak Ridge, Tennessee 37831, USA}
\newcommand{\orsay}{IPN-Orsay, Univ.~Paris-Sud, CNRS/IN2P3, Universit\'e Paris-Saclay, BP1, F-91406, Orsay, France}
\newcommand{\peking}{Peking University, Beijing 100871, People's Republic of China}
\newcommand{\pnpi}{PNPI, Petersburg Nuclear Physics Institute, Gatchina, Leningrad region, 188300, Russia}
\newcommand{\riken}{RIKEN Nishina Center for Accelerator-Based Science, Wako, Saitama 351-0198, Japan}
\newcommand{\rikjrbrc}{RIKEN BNL Research Center, Brookhaven National Laboratory, Upton, New York 11973-5000, USA}
\newcommand{\rikkyo}{Physics Department, Rikkyo University, 3-34-1 Nishi-Ikebukuro, Toshima, Tokyo 171-8501, Japan}
\newcommand{\saispbstu}{Saint Petersburg State Polytechnic University, St.~Petersburg, 195251 Russia}
\newcommand{\seoulnat}{Department of Physics and Astronomy, Seoul National University, Seoul 151-742, Korea}
\newcommand{\stonybrkc}{Chemistry Department, Stony Brook University, SUNY, Stony Brook, New York 11794-3400, USA}
\newcommand{\stonycrkp}{Department of Physics and Astronomy, Stony Brook University, SUNY, Stony Brook, New York 11794-3800, USA}
\newcommand{\tenn}{University of Tennessee, Knoxville, Tennessee 37996, USA}
\newcommand{\titech}{Department of Physics, Tokyo Institute of Technology, Oh-okayama, Meguro, Tokyo 152-8551, Japan}
\newcommand{\tsukuba}{Tomonaga Center for the History of the Universe, University of Tsukuba, Tsukuba, Ibaraki 305, Japan}
\newcommand{\vandy}{Vanderbilt University, Nashville, Tennessee 37235, USA}
\newcommand{\weizmann}{Weizmann Institute, Rehovot 76100, Israel}
\newcommand{\wigner}{Institute for Particle and Nuclear Physics, Wigner Research Centre for Physics, Hungarian Academy of Sciences (Wigner RCP, RMKI) H-1525 Budapest 114, POBox 49, Budapest, Hungary}
\newcommand{\yonsei}{Yonsei University, IPAP, Seoul 120-749, Korea}
\newcommand{\zagreb}{Department of Physics, Faculty of Science, University of Zagreb, Bijeni\v{c}ka c.~32 HR-10002 Zagreb, Croatia}
\affiliation{\abilene}
\affiliation{\augie}
\affiliation{\banaras}
\affiliation{\barc}
\affiliation{\baruch}
\affiliation{\bnlcoll}
\affiliation{\bnlphys}
\affiliation{\caucr}
\affiliation{\charlesczech}
\affiliation{\chonbuk}
\affiliation{\cns}
\affiliation{\colorado}
\affiliation{\columbia}
\affiliation{\czechtech}
\affiliation{\debrecen}
\affiliation{\elte}
\affiliation{\eszterhazy}
\affiliation{\ewha}
\affiliation{\fsu}
\affiliation{\gsu}
\affiliation{\hiroshima}
\affiliation{\howard}
\affiliation{\ihepprot}
\affiliation{\illuiuc}
\affiliation{\inrras}
\affiliation{\instpasczech}
\affiliation{\isu}
\affiliation{\jaea}
\affiliation{\kek}
\affiliation{\korea}
\affiliation{\kurchatov}
\affiliation{\kyoto}
\affiliation{\lawllnl}
\affiliation{\losalamos}
\affiliation{\lund}
\affiliation{\lyon}
\affiliation{\maryland}
\affiliation{\mass}
\affiliation{\michigan}
\affiliation{\muhlenberg}
\affiliation{\nara}
\affiliation{\natmephi}
\affiliation{\newmex}
\affiliation{\nmsu}
\affiliation{\ohio}
\affiliation{\ornl}
\affiliation{\orsay}
\affiliation{\peking}
\affiliation{\pnpi}
\affiliation{\riken}
\affiliation{\rikjrbrc}
\affiliation{\rikkyo}
\affiliation{\saispbstu}
\affiliation{\seoulnat}
\affiliation{\stonybrkc}
\affiliation{\stonycrkp}
\affiliation{\tenn}
\affiliation{\titech}
\affiliation{\tsukuba}
\affiliation{\vandy}
\affiliation{\weizmann}
\affiliation{\wigner}
\affiliation{\yonsei}
\affiliation{\zagreb}
\author{C.~Aidala} \affiliation{\michigan} 
\author{Y.~Akiba} \email[PHENIX Spokesperson: ]{akiba@rcf.rhic.bnl.gov} \affiliation{\riken} \affiliation{\rikjrbrc} 
\author{M.~Alfred} \affiliation{\howard} 
\author{V.~Andrieux} \affiliation{\michigan} 
\author{N.~Apadula} \affiliation{\isu} 
\author{H.~Asano} \affiliation{\kyoto} \affiliation{\riken} 
\author{B.~Azmoun} \affiliation{\bnlphys} 
\author{V.~Babintsev} \affiliation{\ihepprot} 
\author{A.~Bagoly} \affiliation{\elte} 
\author{N.S.~Bandara} \affiliation{\mass} 
% ...
\collaboration{PHENIX Collaboration}  \noaffiliation

\begin{abstract}


We report the transverse single-spin asymmetries of $J/\psi$ production
at forward and backward rapidity, $1.2<|y|<2.2$, as a function of
$J/\psi$ transverse momentum ($p_T$) and Feynman-$x$ ($x_F$). The data
analyzed were recorded by the PHENIX experiment at the Relativistic
Heavy Ion Collider in 2015 from $p$$+$$p$, $p$$+$Al, and $p$$+$Au
collisions with transversely polarized proton beams 
at $\sqrt{s_{_{NN}}}=200$ GeV. At this collision energy, single-spin
asymmetries for heavy-flavor particle production of $p$$+$$p$ collisions
provide access to the spin-dependent gluon distribution and higher-twist
correlation functions inside the nucleon, such as the gluon Qiu-Sterman
and trigluon correlation functions.  Proton+nucleus collisions offer an
excellent opportunity to study nuclear effects on the correlation
functions. The data indicate a positive asymmetry at the two-standard-deviation
level in the $p$$+$$p$ data for 2 GeV/$c<p_T<10$ GeV/$c$ at backward
rapidity, and negative asymmetries at the two-standard-deviation level in the
$p$$+$Au data for $p_T<2$ GeV/$c$ at both forward and backward rapidity, while
in $p$$+$Al collisions the asymmetries are consistent with zero within the
range of experimental uncertainties.

\end{abstract}

\maketitle

\section{INTRODUCTION}

In polarized $p$+$p$ collisions, the transverse single spin asymmetry 
(TSSA), $A_N$, is defined as the amplitude of the azimuthal angular 
modulation of the outgoing particle's scattering cross section with 
respect to the transverse spin direction of the polarized proton. Early 
theoretical predictions which were purely based on perturbative 
calculations showed that the TSSA should be inversely proportional to 
the hard scale of the scattering~\cite{Kane:1978nd}, and if applied to 
the reactions at energies accessible at the Relativistic Heavy Ion 
Collider (RHIC), the asymmetry would be very small, of order 
$10^{-4}$.  However, the experimental asymmetries of light-flavored 
hadrons turned out to be much larger, 
$A_N=O(10^{-1})$~\cite{Klem:1976ui,Aschenauer:2015eha}.

To explain what has been observed in experiments, several 
theoretical 
frameworks~\cite{Sivers:1989cc,Sivers:1990fh,Collins:1992kk,Efremov:1984ip,Qiu:1991pp} 
were developed in the 1990s. In the collinear factorization framework, 
contributions from multi-parton correlations to the 
transverse-spin-dependent cross section were introduced through three 
types of spin-momentum correlations: (1) twist-3 correlation functions 
of a polarized hadron $\phi_{a/A}^{(3)}(x_1,x_2,\vec{s}_\perp)$ 
convolved with leading-twist correlation functions of an unpolarized 
hadron $\phi_{b/B}(x')$ and with leading-twist parton fragmentation 
functions $D_{c\rightarrow C}(z)$, (2) transversity parton distribution 
functions $\delta q_{a/A}(x,\vec{s}_\perp)$ convolved with twist-3 
correlation functions of an unpolarized hadron 
$\phi_{b/B}^{(3)}(x'_1,x'_2)$ and leading-twist parton fragmentation 
functions $D_{c\rightarrow C}(z)$, and (3) transversity parton 
distribution functions $\delta q_{a/A}(x,\vec{s}_\perp)$ convolved with 
leading-twist correlation functions of an unpolarized hadron 
$\phi_{b/B}(x')$ and with twist-3 fragmentation functions 
$D_{c\rightarrow C}^{(3)}(z_1,z_2)$~\cite{Qiu:1991wg}:

\begin{equation}
\begin{aligned}
\label{anprop}
\ A_N \propto &~\sum_{abc}\phi_{a/A}^{(3)}(x_1,x_2,\vec{s}_\perp) \otimes \phi_{b/B}(x^{\prime}) \otimes \hat{\sigma} \otimes D_{c\rightarrow C}(z)+\\
&\sum_{abc}\delta q_{a/A}(x,\vec{s}_\perp) \otimes \phi^{(3)}_{b/B}(x_1^{\prime},x_2^{\prime}) \otimes \hat{\sigma}^{\prime} \otimes D_{c\rightarrow C}(z)+\\
&\sum_{abc}\delta q_{a/A}(x,\vec{s}_\perp) \otimes \phi_{b/B}(x^{\prime}) \otimes \hat{\sigma}^{\prime\prime} \otimes D^{(3)}_{c\rightarrow C}(z_1,z_2).
\end{aligned}
\end{equation}
In this notation, $a/A$ means the distribution of parton $a$ in hadron 
$A$, $b/B$ means the distribution of parton $b$ in hadron $B$ and 
$c\!\rightarrow \!C$ means the fragmentation of parton $c$ into hadron 
$C$. Additionally, $x$ is the Bjorken parton momentum fraction of the 
incoming hadron; $z$ is the fraction of the outgoing partonic momentum 
carried by the detected hadron; $\vec{s}_\perp$ is the transverse spin 
of the incoming hadron; $\hat{\sigma}$, $\hat{\sigma}^{\prime}$, and 
$\hat{\sigma}^{\prime\prime}$ are the partonic hard-scattering cross 
sections of a process where higher twist is associated with the incoming 
polarized or unpolarized hadron, or outgoing partons, respectively. 
Non-zero values for $\phi_{a/A}^{(3)}(x_1,x_2,\vec{s}_\perp)$ and 
$D_{c\rightarrow C}^{(3)}(z_1,z_2)$ in Eq.~\ref{anprop} account for the 
large $A_N$ observed in experiment, where 
$\phi_{a/A}^{(3)}(x_1,x_2,\vec{s}_\perp)$ corresponds to initial-state 
effects~\cite{Qiu:1991pp} and $D_{c\rightarrow C}^{(3)}(z_1,z_2)$ to 
final-state effects~\cite{Kang:2010zzb}. Initial-state effects are 
described by the twist-3 three-parton correlation functions 
$\phi_{a/A}^{(3)}(x_1,x_2,\vec{s}_\perp)$ and 
$\phi_{b/B}^{(3)}(x'_1,x'_2)$ which measure the quantum interference 
between two scattering amplitudes of the incoming 
hadron~\cite{Kang:2008ey}, while final-state effects are related to 
twist-3 fragmentation functions $D_{c\rightarrow C}^{(3)}(z_1,z_2)$ 
which describe the process in which the outgoing parton fragments into a 
final-state hadron~\cite{Kang:2008ey}. At RHIC energies, heavy quark 
production, such as $J/\psi$ production, is dominated by gluon-gluon 
interactions. Because the gluon transversity distribution does not 
exist, the second and third terms of Eq.~\ref{anprop} are zero. This 
means that heavy flavor $A_N$ is free from final-state effects and is 
sensitive to initial-state effects, such as the gluon Qiu-Sterman and 
trigluon correlations which correspond to the factor 
$\phi_{a/A}^{(3)}(x_1,x_2,\vec{s}_\perp)$ in 
Eq.~\ref{anprop}~\cite{Kang:2008qh}.

In the case of high energy hadronic collisions, a nonvanishing SSA is 
generated by a parton-level spin flip and a phase difference between the 
scattering amplitude and the corresponding complex conjugate. In the 
context of the quantum chromodynamic (QCD) collinear-factorization 
framework, the parton-level spin flip is generated by the interference 
between the active single parton and a two-parton composite state of the 
scattering amplitude. On the other hand, the phase difference is 
achieved by the interference between the real and imaginary part of the 
partonic scattering 
amplitude~\cite{Qiu:1991pp,Qiu:1991wg,PhysRevD.59.014004}. For the 
Qiu-Sterman correlation, the quantum interference is between a quark 
state of momentum fraction $x$ and a quark-gluon composite state with 
the same momentum fraction where either the gluon or quark carries the 
total momentum of the quark-gluon composite state~\cite{Kang:2010hg}. In 
the trigluon correlation, the two parton composite state is composed of 
two gluons instead of a quark and gluon as described above for the 
Qiu-Sterman correlation~\cite{Kang:2008ey,Braun:2009mi}. The collinear 
factorization framework has been widely used to describe the TSSA's 
measured at 
RHIC~\cite{Adams:2003fx,Abelev:2008af,Adamczyk:2012xd,Adler:2005in,Adare:2014qzo,Adare:2013ekj,Aidala:2017pum,Arsene:2008aa}.

An alternative treatment is known as the Transverse-Momentum-Dependent 
(TMD) formalism. In this formalism, the cross section is factorized into 
hard-scattering cross sections and TMD parton distribution and 
fragmentation functions (PDFs and FFs)~\cite{Anselmino:2012rq}. For the 
TMD approach to be valid in the context of $p$+$p$ collisions, $Q^2$ 
must be large, in order to use perturbative QCD, while the transverse 
momentum must satisfy $p_T\ll Q$ and not be much larger than the 
intrinsic, parton transverse momentum $k_T$, so that effects of the 
latter remain visible~\cite{Collins:1984kg}. One of the TMD PDFs, called 
the Sivers function~\cite{Sivers:1989cc}, is widely used in describing 
the TSSA's that were observed in different 
processes~\cite{Airapetian:2004tw,Adolph:2017pgv,Godbole:2016ixc,DAlesio:2015fwo,Mukherjee:2016qxa}. 
The Sivers function, denoted by 
$f^{\perp}_{1T}(x,\boldsymbol{k}_{\perp}^{2})$, describes the distortion 
in the distribution of unpolarized partons with momentum fraction $x$ 
and transverse momentum $\boldsymbol{k}_{\perp}$ in a transversely 
polarized hadron. This distorted distribution of unpolarized partons 
causes an azimuthal anisotropy in the distribution of parton transverse 
momenta in the polarized hadron which gives rise to the nonzero TSSA. 
As it has been described above, at low $p_T$, the nonperturbative TMD 
Sivers function will be responsible for its SSA, while twist-3 dominates 
the contributions to the SSA when $p_T \sim Q$. At intermediate $p_T$, 
one can see the transition between these two frameworks and a relation 
between Sivers Function and Qiu-Sterman Function has been shown in 
Ref.~\cite{Boer:2003cm}.

With increasing experimental information on the quark Sivers function 
during the last ten years, our understanding of this quantity has 
matured~\cite{Anselmino:2013rya,Anselmino:2008sga,Collins:2005ie,Vogelsang:2005cs,Anselmino:2005ea}, 
while the gluon Sivers function is still relatively unknown. The 
transversely polarized $p$$+$$p$ collisions studied at RHIC present a very 
good opportunity to study the gluon Sivers function as gluon-gluon 
interactions are dominant in $p$$+$$p$ collisions at RHIC energies. PHENIX 
has measured the TSSA for $J/\psi$ production at central and forward 
rapidities~\cite{Adare:2010bd} and, at small $p_T$ values, where the 
$J/\psi$ mass becomes the large scale $Q$ in TMD factorization, the 
result has been compared to a gluon Sivers function derived in the 
context of the color-evaporation model in~\cite{Godbole:2017syo} 
and the generalized-parton model in~\cite{DAlesio:2017rzj}.

In proton-nucleus ($p$+$A$) collisions, the increase of the atomic 
number results in increasing gluon occupancy and therefore gluon 
saturation effects may become important in the small $x$ region. In the 
Regge-Gribov limit, the properties of saturated gluons in the 
infinite-momentum frame can be described by the Color Glass Condensate 
(CGC) which has been applied to a variety of processes such as 
$e$$+$$p$, $e$$+$$A$, $A$$+$$A$, and $p$$+$$A$ 
collisions~\cite{Gelis:2010nm}. The quark and gluon distribution 
functions for large nuclei were computed first in 
Ref.~\cite{McLerran:1993ni} in the weak coupling limit. Using the CGC 
framework, one can describe the rescatterings of the outgoing parton 
within the nucleus. In the coherent QCD multiple scattering 
framework~\cite{Qiu:2004da}, it has been shown that at low $p_T$ the 
rates of single and double hadron production are highly suppressed and 
the amount of suppression grows with rapidity and centrality; meanwhile, 
at high $p_T$, such nuclear modification effects become less pronounced 
as we enter the perturbative region in a dilute nuclear medium. For the 
computation of TSSA's, a hybrid approach has been widely used in photon, 
$\gamma$-jet and dijet 
production~\cite{Schafer:2014zea,Schafer:2014xpa,Zhou:2017sdx,Zhou:2017mpw}. 
The hybrid approach treats the gluon distribution inside the heavy 
nucleus in the CGC framework and utilizes the twist-3 formalism for the 
proton side.The resummation of the power corrections in $p$$+A$ 
reactions shifts the nuclear PDF to higher $x$~\cite{Qiu:2004da}. The 
new PHENIX $p$$+A$ collision data offer the opportunity to quantify how 
this shift in $x$ affects the twist-3 description of the TSSA's 
discussed above.

\section{EXPERIMENT SETUP}

\begin{figure*}[thb]
\begin{minipage}{0.8\linewidth}
\includegraphics[width=0.99\linewidth]{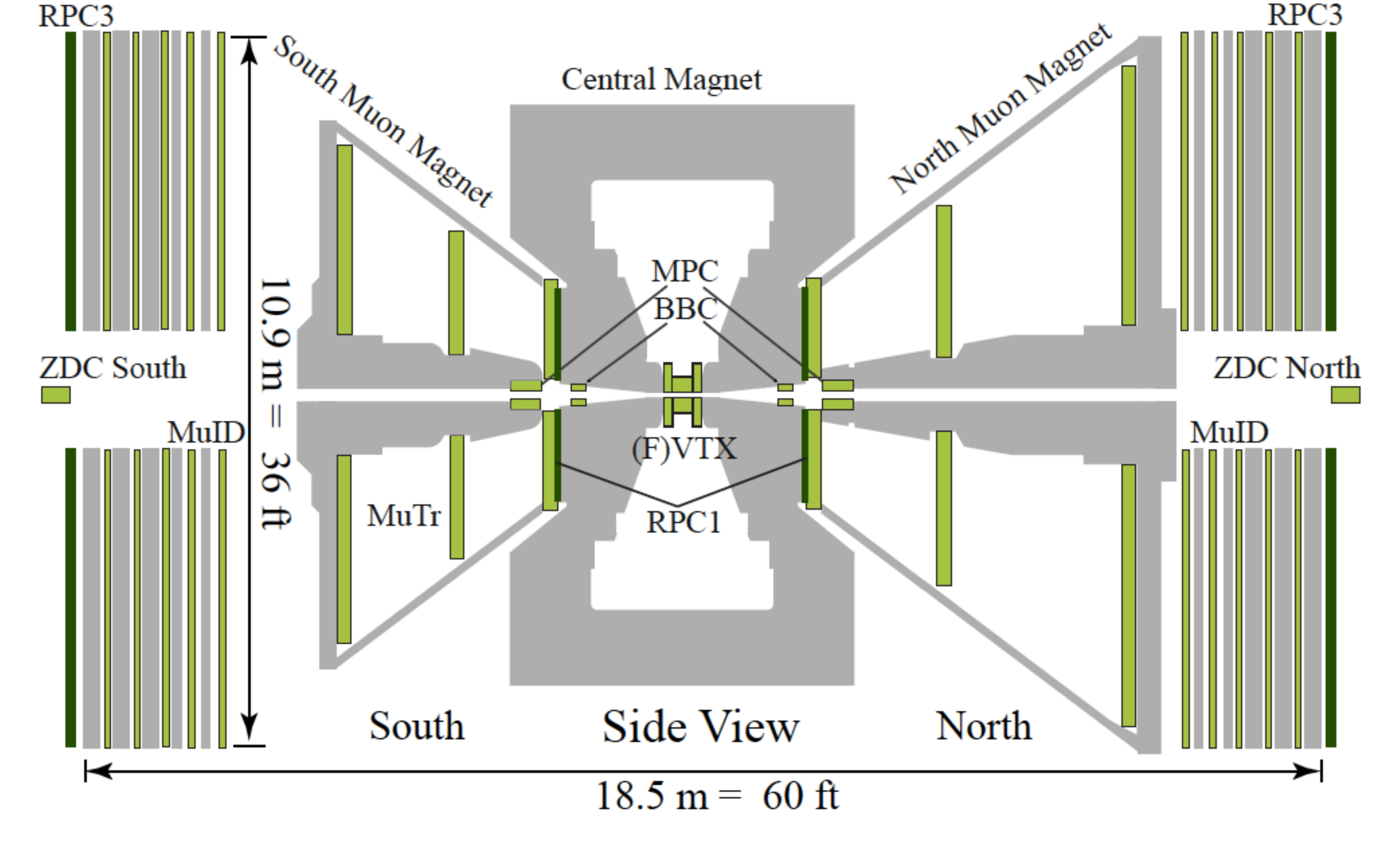}
\end{minipage}
\begin{minipage}{0.18\linewidth}
\caption{\label{fig:phenix}Side view of the 2015 PHENIX detector. Of 
primary importance to this analysis are the BBC, MuTr, and MuID.  
See text for descriptions of these subsystems and how they were used.}
\label{fig:PHENIX}
\end{minipage}
\end{figure*}

The $J/\psi$ mesons are measured by detecting muon pairs 
$\mu^{+}\mu^{-}$ in the two PHENIX muon spectrometers. The two muon 
spectrometers with full azimuthal coverage cover a range in 
pseudorapidity $\eta\in{[-2.2,-1.2]}$ for the south arm and 
$\eta\in{[1.2,2.4]}$ for the north arm (see Fig.~\ref{fig:PHENIX}). The 
momenta of the muons are measured by the muon trackers (MuTr). The MuTr 
is composed of three stations of cathode-strip tracking chambers inside 
a radial magnetic field~\cite{PHENIX:2003}. The MuTr chambers are 
followed by the muon identifiers (MuID) which contain 5 sensitive layers 
(named gap0--gap4) per arm, with each layer made of one vertically and 
one horizontally oriented Iarocci tube plane, all interleaved with 10- 
or 20-cm thick steel absorber plates to suppress hadron backgrounds. 
With the front absorber provided by the PHENIX central magnets and the 
MuID together with additional stainless steel absorber mounted in 2011, 
the total thickness of the steel absorbers is about 190 cm. The 
probability for a 4 GeV/$c$ pion to punch through the whole absorber is 
less than 3\%~\cite{Adcox:2003zm,Adachi:2013qha}. Muons with momenta of 
at least 3 GeV/$c$ can penetrate all the absorbers and reach the last 
layer of the MuID with high efficiency.

Two beam-beam counters (BBC) are located at opposite ends at 144~cm from 
the interaction point along the beam line with full azimuthal coverage 
and $|\eta| \in[3.1,3.9]$. Each BBC has 64 elements consisting of quartz 
\v{C}erenkov radiator and mesh-dynode PMT. The BBCs also serve as an 
interaction vertex finder with resolution along the beam direction of 
about 2~cm and the z-vertex approximately follows a Gaussian distribution centered at 0 with 
a width of about 10 cm in $p$+$p$ collisions and in addition play the role of a 
luminosity detector~\cite{PHENIX_FULL:2003}. For the minimum-bias (MB) 
trigger, it is required to have one or more hits in each BBC.  The 
MB-trigger efficiency for $p$$+$$p$ collisions is about 
55\%~\cite{Adler:2003pb}, while this trigger is 84\% (72\%) efficient 
for $p$$+$Au ($p$$+$Al) collisions; the percentage is defined with 
multiplicity in the south BBC (A-going direction)~\cite{Adare:2013nff}.

Events containing dimuon candidates were selected using the 
combination of BBC-MB trigger and other muon triggers. The 
``2-deep muon trigger" requires that both muon tracks have at 
least one hit in the last two MuID gaps and no less than two 
hits in other gaps. The ``sagitta-3 muon trigger" selects 
tracks that were recorded by MuTr with high momentum, 
requiring that the maximum track sagitta, determined by the 3 
MuTr stations, be less than 3 MuTr cathode strips at the 
middle plane.

The dimuon candidates are then screened by the dimuon cuts 
shown in Table~\ref{tab:cuts}. Here, $p_z$ and $p_T$ are 
respectively the longitudinal and transverse momentum of the 
dimuon pair with respect to the beam direction, the lower-side 
$p_T$ cut is imposed due to the MuTr resolution, DG0 is the 
distance between two matched tracks coming one each from the 
MuTr and MuID, projected to the first MuID gap, DDG0 is the 
opening angle between these projected tracks, ntrhits 
(nidhits) is the number of hits generated by a track in the 
MuTr (MuID), lastgap is the last-gap number in the MuID that a 
track penetrates, and DCA$_{r}$ is the distance of closest 
approach between a muon track and the beam line. 
The event vertex determined by the BBC follows a Gaussian
distribution centered at 0 with a width of about 40 cm.
A z-position vertex cut within $\pm 30$ cm ensures
that the collision occurs inside the spectrometer acceptance
and keeps approximately 50\% of all collisions.
The vertex $\chi^{2}$ is determined by 
fitting the two candidate tracks with the event vertex given 
by the BBC.

\begin{table}[tbh]
\caption{\label{tab:cuts}
Track selections and cuts used in this analysis.  See text for
definitions and details.}
\begin{ruledtabular} \begin{tabular}{cccc}
 & Single muon cuts 
 & DG0 $<30$~cm (south)    &\\
&& DG0 $<25$~cm (north)    &\\
&& DDG0 $<10^\circ$        &\\
&& lastgap $>2$            &\\
&& ntrhits $>9$            &\\
&& nidhits $>5$            &\\
&& MuTr DCA$_{r}$ $<10$~cm &\\
\\
 & Dimuon cuts
 & $p_{z} <$ 100 GeV/$c$&\\
&& 0.42 GeV/$c$ $< p_{T} <$ 10 GeV/$c$&\\
&& -2.2 $< \eta <$ -1.2 (south)&\\
&& 1.2 $< \eta <$ 2.2 (north) &\\
&& vertex $\chi^{2} < 5$&\\
&& $-30$ cm $<{\rm BBC}_{z}$ vertex $<30$ cm &\\
\end{tabular} \end{ruledtabular}
\end{table}

\begin{table}[bht]
\caption{\label{tab:polarization}Blue and yellow beam polarization in 
$p$$+$$p$ and $p$$+A$ collisions.}
\begin{ruledtabular} \begin{tabular}{ccccc}
& Data set    & Blue beam & Yellow beam  &\\
\hline
& $p$$+$$p$   & 57\%      & 57\%         &\\
& $p$$+$Al    & 57\%      & 0            &\\
& $p$$+$Au    & 61\%      & 0            &\\ 
\end{tabular} \end{ruledtabular}
\end{table}


In the 2015 RHIC run, we took 11 weeks of $p$+$p$ collision data with an 
integrated luminosity of about 40 pb$^{-1}$ and with both the 
blue beam (clockwise) and yellow beam 
(counter-clockwise) transversely (vertically) polarized at the PHENIX 
interaction point. In $p$$+A$ collisions, only the blue beam (the 
proton beam) was polarized; we have 2- and 5-week data sets for 
$p$$+$Al and $p$$+$Au with integrated luminosities of 
about 6.0~pb$^{-1}$ and 6.6~pb$^{-1}$, respectively. 
Table~\ref{tab:polarization} shows the beam polarizations.
Each proton 
beam was filled into 111 bunches with different spin states in 8 base 
patterns~\cite{Alekseev:1489159}.

\section{DATA ANALYSIS}

The maximum likelihood method was used to extract the transverse single 
spin asymmetry $A_N$. The likelihood $\mathcal{L}$ for one dimuon pair 
with azimuthal angle $\phi$ with respect to the incoming polarized 
proton beam direction is 1 + $P \cdot A_{N}\sin(\phi_{\rm pol} - 
\phi)$, where $P$ is the beam polarization, and $\phi_{\rm pol}$ is 
the direction of beam polarization which is $+(-)$$ \pi/2$ when the spin 
is up (down). Then the log-likelihood for all the dimuon pairs is given 
by:

\begin{equation}
\label{eq:maxlikelihood}
\log \mathcal{L} = \sum_{i} \log(1 + P \cdot A_N \sin(\phi_{\rm pol} - \phi_i)).
\end{equation}

We select the value of $A_N$ that maximizes $\mathcal{L}$. The 
statistical uncertainty of $A_N$ is obtained by calculating the inverse 
of the second derivative of $\mathcal{L}$ with respect to $A_N$:

\begin{equation}
\label{eq:maxlikelihood_err}
\sigma^2(A_N) = (-\frac{\partial^2 \log\mathcal{L}}{\partial A_N^2})^{-1}.
\end{equation}

\begin{figure}[thb]
\includegraphics[width=0.97\linewidth]{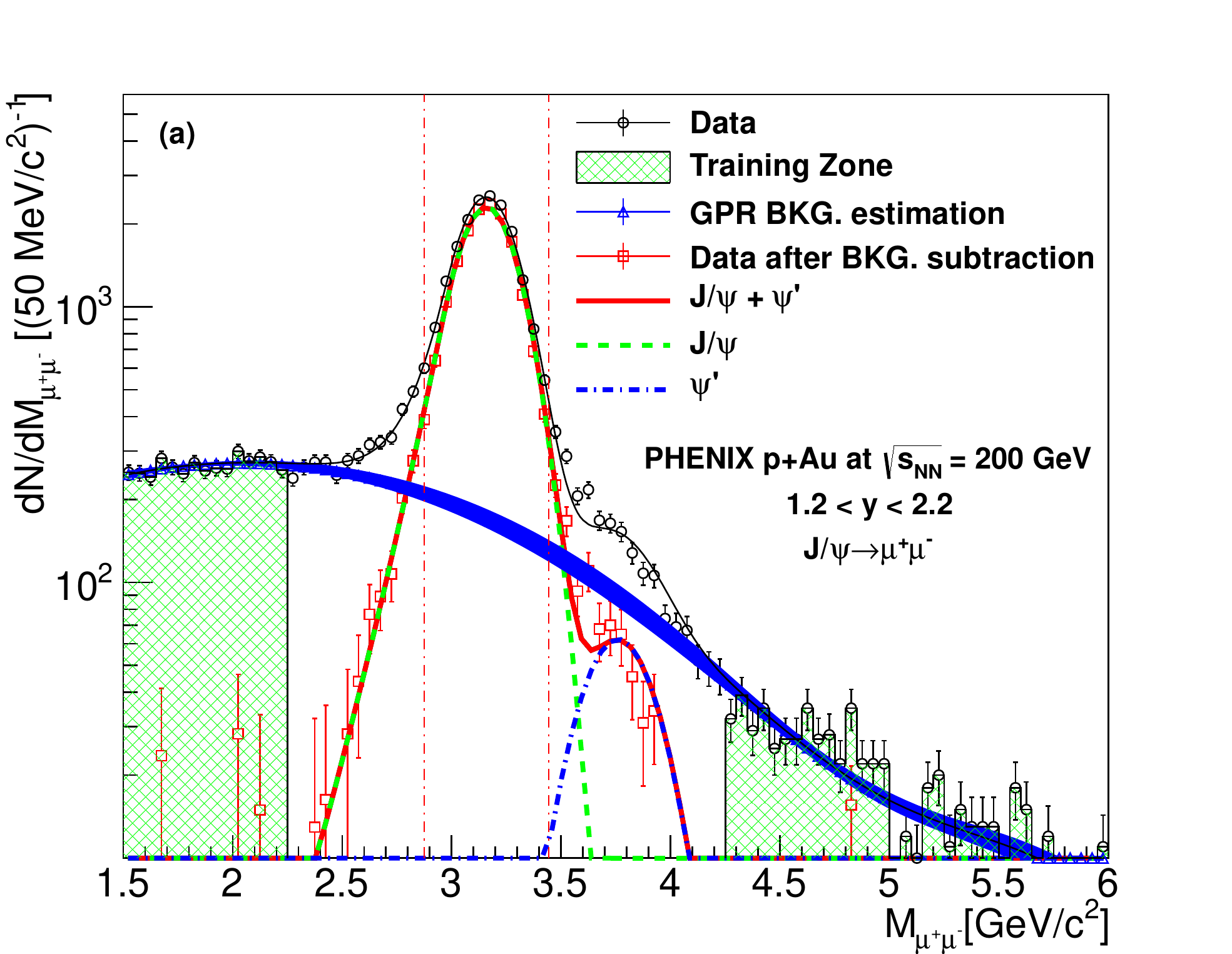}
\includegraphics[width=0.97\linewidth]{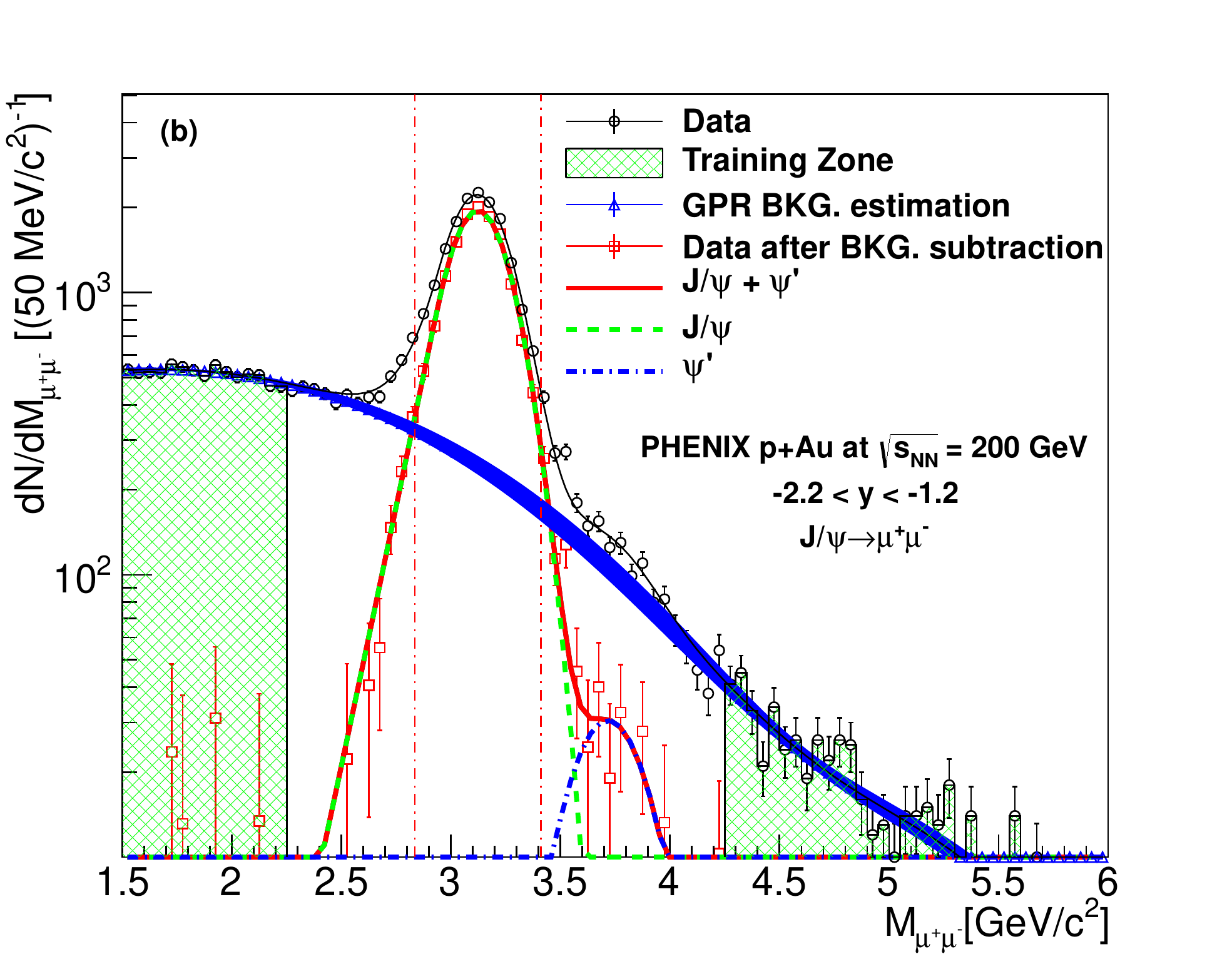}
\caption{\label{fig:example}Dimuon mass spectrum fits with GPR 
background estimation for $0.42<p_T<10$~GeV/$c$ for the 
(a) $p$-going and (b) Au-going directions.}
\end{figure}

\begin{figure}[thb]
\includegraphics[width=1.0\linewidth]{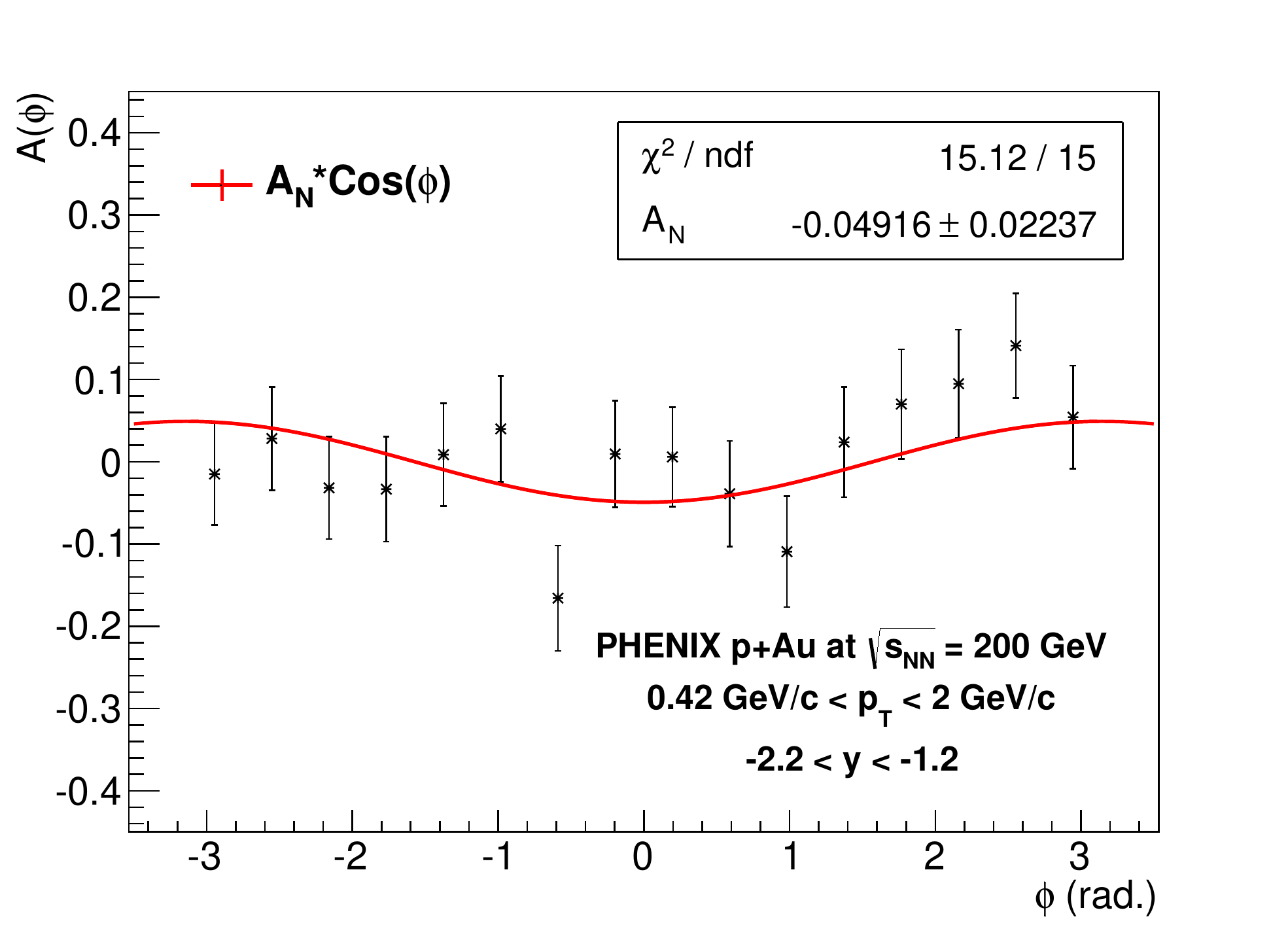}
\caption{\label{fig:az_fit} Fit of inclusive A($\phi$) with a cosine 
modulation. The asymmetry and $\chi^2$ per degree of freedom from the 
fit are shown.}
\end{figure}

To extract the $J/\psi$ $A_N$, first we calculated the inclusive $A_N$ 
($A_N^{\rm incl}$) in the $J/\psi$ invariant mass region between 2.8 
GeV/$c^2$ and 3.4 GeV/$c^2$. Then we corrected that $A_N$ with the 
background $A_N$ ($A_N^{\rm bgr}$) using Eq.~\ref{aneq} and 
calculated the corresponding uncertainty in $A_N^{J/\psi}$ using 
Eq.~\ref{anerreq}:

\begin{equation}
\label{aneq}
A_N^{J/\psi}=\frac{A_N^{\rm incl}-f\cdot A_N^{\rm bgr}}{1-f}
\end{equation}

\begin{equation}
\label{anerreq}
\delta A_N^{J/\psi}=\frac{\sqrt{(\delta
A_N^{\rm incl})^2+f^2\cdot (\delta A_N^{\rm bgr})^2}}
{1-f}.
\end{equation}

\noindent
Here, $f$ is the background fraction in the the $J/\psi$ 
invariant mass region. The statistical uncertainty from $f$ is 
not taken into account because its contribution to the total 
statistical uncertainty is small ($< 5\%$).

The background $A_N$ under the $J/\psi$ peak was estimated 
from a sideband in the dimuon invariant mass range of 1.5-2.4 
GeV/$c^2$; a similar approach was used in previous PHENIX 
measurements~\cite{Adare:2015ozj,Adare:2016cqe}. The 
background fraction $f$ was obtained using the Gaussian 
Process Regression (GPR) 
method~\cite{MacKay,Rasmussen,Lauritzen,barber2012bayesian} 
for each $p_T$ and $x_F$ bin. The GPR method is a 
nonparametric regression approach considered to be less 
biased, and has been used in several previous PHENIX 
measurements~\cite{Adare:2015gsd,Adare:2016cqe}. 
Figure~\ref{fig:example} shows the fitting result with the GPR 
method using $p$+\rm{Au} data. The dimuon invariant mass 
windows of 1.5 GeV/$c^2$ to 2.2 GeV/$c^2$ and 4.3 GeV/$c^2$ to 
6.0 GeV/$c^2$ are used for estimating the background shape 
(see Fig.~\ref{fig:example}). Then the signal ($J/\psi$ and 
$\psi(2S)$) yields are obtained by subtracting the GPR 
background from the inclusive dimuon spectrum. The $J/\psi$ 
and $\psi(2S)$ peaks are fitted respectively by a Crystal Ball 
function~\cite{Skwarnicki:1986xj} and a Gaussian function. The 
vertical lines in Fig.~\ref{fig:example} represent $\pm 
2\sigma$ mass windows around the $J/\psi$ peak, where the 
$\psi(2S)$ contribution will be negligible; $J/\psi$ 
production is dominant in both the $p$-going and Au-going 
directions. The background fraction is 17\% for the Au-going 
direction, 13\% for the Al-going direction and 10\% for the 
$p$-going direction.

The $A_N^{\rm incl}$ and $A_N^{\rm bgr}$ asymmetries for each 
$p_T$ and $x_F$ bin are calculated with the maximum likelihood 
method described above but with different dimuon invariant 
mass windows. For $A_N^{\rm incl}$ we used unlike-sign muon 
pairs in the invariant mass range $\pm2\sigma$ around the 
$J/\psi$ peak, while for $A_N^{bgr}$ we used the fixed 
invariant mass range from 1.5 GeV/$c^2$ to 2.4 GeV/$c^2$.

\begin{figure*}[thb]
\includegraphics[width=0.99\linewidth]{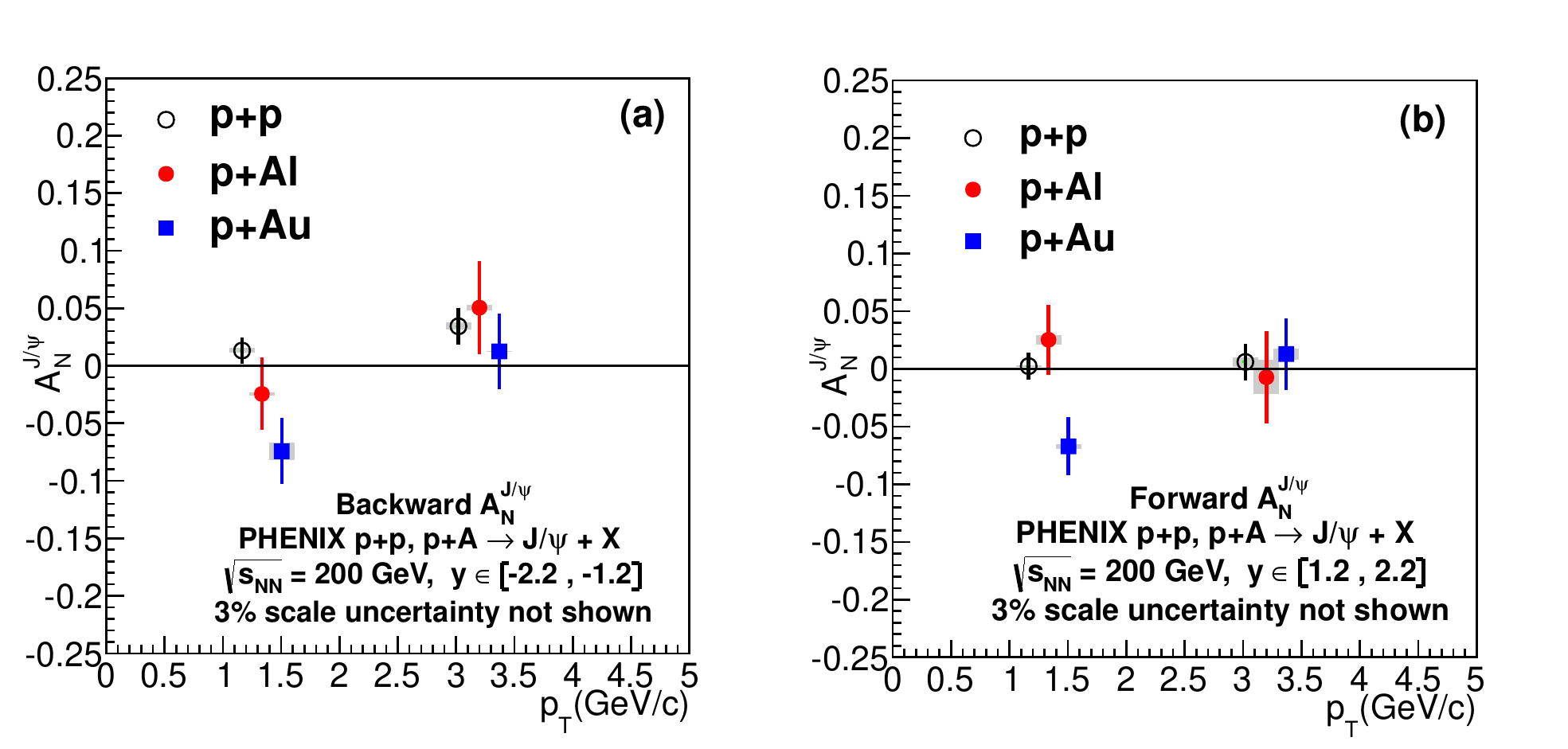}
\caption{\label{fig:final_pt} (a) Backward [$x_F<0$] and (b) forward 
[$x_F>0$] $A_N^{J/\psi}$ vs $p_T$ 
for open [black] circles $p$$+$$p$, 
closed [red] circles $p$$+$Al, and 
closed [blue] boxes $p$$+$Au collisions.  
The shaded [gray] boxes show the systematic uncertainty.  
The data points for $p$$+$Al and $p$$+$Au collisions 
have been shifted in $p_T$ for clarity.}
\end{figure*}

In this analysis, there are two sources of systematic 
uncertainty, detailed as sources 1 and 2 in the following two 
paragraphs and listed quantitatively in Tables III and IV. For 
the total systematic uncertainty displayed in Figs. 4 and 5, 
we have combined these two sources quadratically.

\begin{figure}[htbp]
\includegraphics[width=1.05\linewidth]{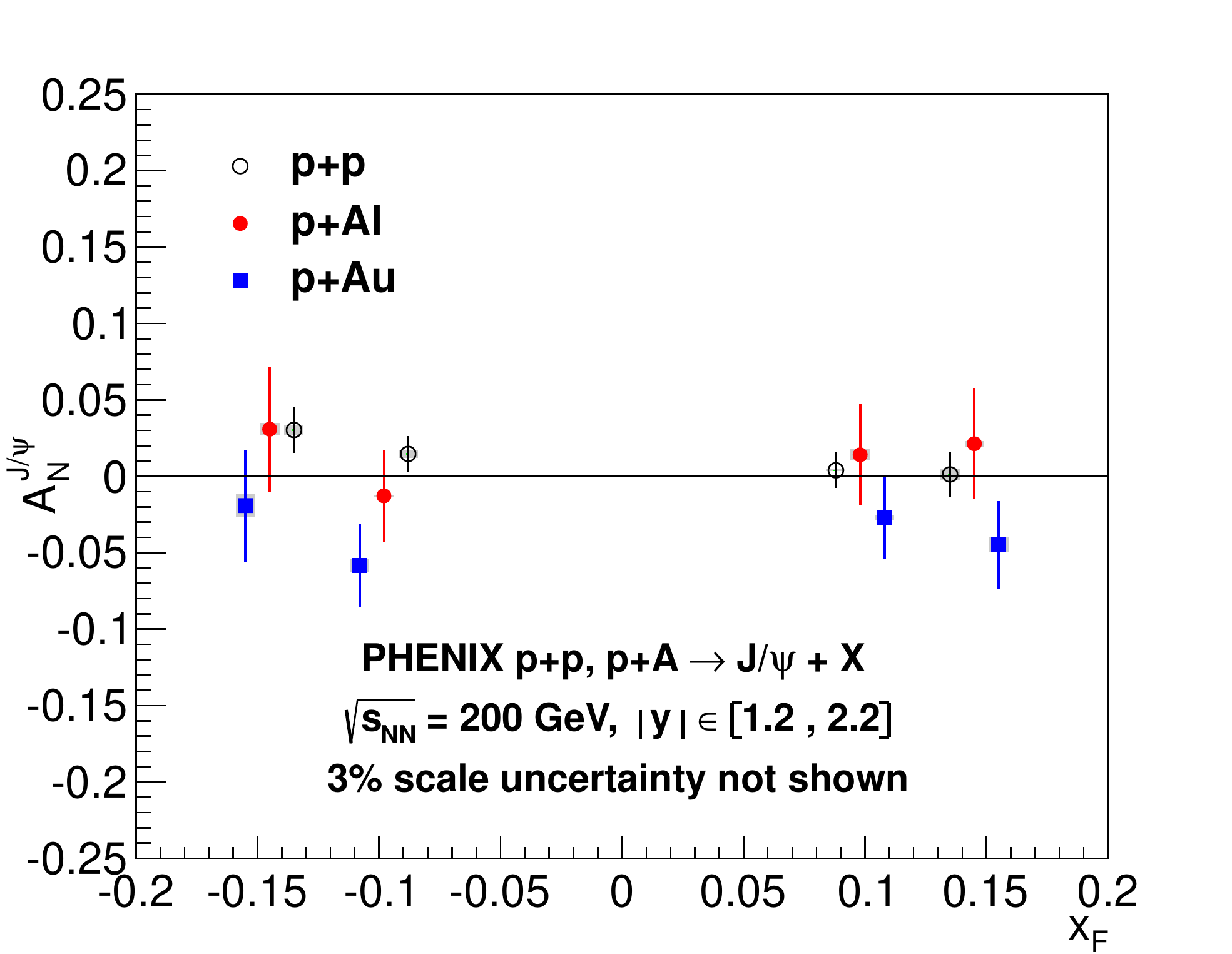}
\caption{\label{fig:final_xf} $A_N^{J/\psi}$ vs $x_F$ for 
for open [black] circles $p$$+$$p$, 
closed [red] circles $p$$+$Al, and 
closed [blue] boxes $p$$+$Au collisions.  
The shaded [gray] boxes show  the systematic uncertainty. 
The data points for $p$$+$Al and $p$$+$Au collisions 
have been shifted in $x_f$ for clarity.}
\end{figure}

The first systematic uncertainty source (``source 1'') concerns the 
method of determining the asymmetry itself.  We check this by 
determining $A_N$ with a different method, the azimuthal fitting method. 
Similar to Eq.~\ref{eq:maxlikelihood}, the production cross section of 
$J/\psi$ as a function of the azimuthal angle $\phi$ is given by:

\begin{equation}
\sigma(\phi) \propto 1 + P \cdot A_N \sin(\phi_{\rm pol} - \phi),
\end{equation}
where $\phi_{\rm pol} = +(-)\pi/2$ when the spin is up (down).

\begin{table*}[thb]
\caption{\label{tab:AN_VS_PT}$A_{N}^{J/\psi}$ for ranges of $p_T$ in 
forward and backward rapidity for $p$$+$$p$, $p$$+$Al, and $p$$+$Au 
collisions.}
\begin{ruledtabular} \begin{tabular}{ccccccccc}
Range of $x_F$ &
$p_{T} ({\rm GeV}/c)$ & Data sample & $\left<p_T\right> ({\rm GeV}/c) $ & $A_N$ 
& $\delta A_{N}^{\rm stat}$ & $\delta A_{N}^{\rm syst}({\rm source~1})$ 
& $\delta A_{N}^{\rm syst}({\rm source~2})$\\
\hline
Forward ($x_F>0$)
&$0.42<p_{T}<2$ & $p$+$p$ 		& 1.17& 0.002& 0.011&0.0006&0.0001\\
& 				& $p$+$\rm{Al}$ & 1.19& 0.025&  0.030    &0.0034&0.0026\\
&                & $p$+$\rm{Au}$ & 1.20& -0.067& 0.025     &0.0008&0.0020\\
\\
&$2<p_{T}<10$ & $p$+$p$ 		& 3.02& 0.006& 0.016 &0.0009&0.0041\\
&				& $p$+$\rm{Al}$ & 3.07& -0.007&0.040      &0.0148&0.0005\\
&                & $p$+$\rm{Au}$ & 3.13&0.013 &0.031      &0.0045&0.0015\\
\\                
Backward ($x_F<0$)
&$0.42<p_{T}<2$ & $p$+$p$ 		& 1.16& 0.013&0.011	   &0.0021&0.0002\\
& 				& $p$+$\rm{Al}$ & 1.18& -0.024&0.031   &0.0007&0.0012\\
&                & $p$+$\rm{Au}$ & 1.19& -0.074&0.029	   &0.0077&0.0008\\
\\
&$2<p_{T}<10$ & $p$+$p$ 		& 3.00& 0.034&0.016	   &0.0027&0.0015\\
&				& $p$+$\rm{Al}$ & 3.03& 0.050&	0.041   &0.0024&0.0001\\
&                & $p$+$\rm{Au}$ & 3.03& 0.013&0.033	   &0.0004&0.0004\\
\end{tabular} \end{ruledtabular}
\caption{\label{tab:AN_VS_XF}$A_{N}^{J/\psi}$ for ranges of $x_F$ in 
forward and backward rapidity for $p$+$p$, $p$$+$Al and $p$$+$Au 
collisions.}
\begin{ruledtabular} \begin{tabular}{ccccccc}
Range of $x_F$                 & Data sample     & $<x_F> $ & $A_N$ & $\delta A_{N}^{stat}$ & $\delta A_{N}^{syst}({\rm source~1})$ & $\delta A_{N}^{syst}({\rm source~2})$\\\hline
$0.05<x_{F}<0.11$ & $p$+$p$ 		& 0.088     & 0.004 & 0.012     &0.0001&0.0006\\
 				& $p$+$\rm{Al}$ 		& 0.089     & 0.014	& 0.033		&0.0029&0.0021\\
                & $p$+$\rm{Au}$ 		& 0.089 	&-0.027	& 0.027		&0.0014&0.0001\\
\\
$0.11<x_{F}<0.30$ & $p$+$p$ 		& 0.135 	&0.001	&0.015		&0.0018&0.0030\\
				& $p$+$\rm{Al}$ 		& 0.135 	&0.021	&0.036		&0.0011&0.0013\\
                & $p$+$\rm{Au}$ 		& 0.136 	&-0.045	&0.029		&0.0049&0.0003\\
\\
$-0.11<x_{F}<-0.05$ & $p$+$p$ 		& -0.086    & 0.013 & 0.012     &0.0026&0.0002\\
 				& $p$+$\rm{Al}$ 		& -0.086    &-0.013	&0.030		&0.0007&0.0004\\
                & $p$+$\rm{Au}$ 		& -0.086  	&-0.058	&0.027		&0.0040&0.0007\\
\\
$-0.30<x_{F}<-0.11$ & $p$+$p$ 		& -0.132  	&0.030	&0.015		&0.0031&0.0011\\
				& $p$+$\rm{Al}$ 		& -0.132  	&0.031	&0.041		&0.0002&0.0041\\
                & $p$+$\rm{Au}$ 		& -0.132  	&-0.019	&0.037   &0.0077&0.0006\\
\end{tabular} \end{ruledtabular}
\end{table*}

The asymmetry can be written as function of azimuthal angle $\phi$ as:
\begin{equation}
\label{avsphi}
\ A(\phi)=\frac{\sigma^{\uparrow}(\phi)-\sigma^{\downarrow}(\phi)}{\sigma^{\uparrow}(\phi)+\sigma^{\downarrow}(\phi)}=A_N\cdot\cos(\phi).
\end{equation}
Therefore, $A_N$ can be extracted by fitting the $A(\phi)$ with a cosine 
modulation. As an example, Figure~\ref{fig:az_fit} shows the 
determination of $A_N^{\rm incl}$ for dimuons with $0.42<p_T<2~{\rm 
GeV}/c$ in the Au-going direction. The differences of $J/\psi$ $A_{N}$ 
determined from the maximum likelihood method and azimuthal fitting 
method are treated as a systematic uncertainty. The value of the 
source 1 uncertainty ranges from 1\% to 35\% of the statistical 
uncertainties.

A second source of systematic uncertainty (``source 2'') is from the 
method of determining the background fraction $f$. We studied a 
potential bias of the GPR method by parameterizing the background with a 
3rd order polynomial instead. The $f$ changed by about 2\% and the 
corresponding difference in the resulting background corrected $J/\psi$ 
$A_{N}$s has been assigned as a systematic uncertainty which is of the 
order of 10\% of the statistical uncertainty.

\section{Results}

Figure~\ref{fig:final_pt} and Table~\ref{tab:AN_VS_PT} show the TSSA for 
$J/\psi$ production, $A_N^{J/\psi}$, in two $p_T$ bins in forward and 
backward kinematics in $p$$+$$p$, $p$$+$Al and $p$$+$Au 
collisions. The 2015 $p$$+$$p$ data are consistent with the previous results 
of $A_N^{J/\psi}$ from the 2006 and 2008 data~\cite{Adare:2010bd} within 
one-standard-deviation. The 2015 $p$$+$$p$ data favor a positive asymmetry 
(at the 2$\sigma$ level) in the high-$p_{T}$ bin in backward rapidity. 
With limited statistics, the $A_N$ in all $p_T$ and $x_F$ bins for 
$p$$+$Al collisions are consistent with zero. In $p$$+$Au 
collisions, the asymmetry in the high-$p_T$ bin is consistent with zero, 
although there is a trend to a nonzero $A_N$ (at the 2$\sigma$ level) 
in the low-$p_T$ bin in both the forward and backward directions.

Figure~\ref{fig:final_xf} and Table~\ref{tab:AN_VS_XF} show the 
$A_N^{J/\psi}$ as a function of $x_F$.  In the $p$$+$$p$ data, it is 
consistent with the previous PHENIX results~\cite{Adare:2010bd} and a 
$\sim2 \sigma$ positive $A_N$ is observed in the backward higher $x_F$ 
bin. The result for the other $x_F$ bins are consistent with zero. For 
the $p$$+$Au data, a $\sim2\sigma$ negative $A_N$ is observed in 
the forward high-$x_F$ bin and the backward low-$x_F$ bin. A scale 
uncertainty from the polarization (3\%) is not included in both 
Figure~\ref{fig:final_pt} and Figure~\ref{fig:final_xf}.

\section{Conclusion}

We have reported the measurements of the transverse single-spin 
asymmetry in $J/\psi$ production at forward and backward rapidity with 
$1.2<|y|<2.2$ in $p$+$p$, $p$+$\rm{Al}$ and $p$+$\rm{Au}$ collisions 
with transversely polarized proton beams at $\sqrt{s_{_{NN}}} = 200$ GeV 
using the RHIC Run 2015 data. The results from $p$$+$$p$ collisions are 
consistent with previous PHENIX results. Within experimental 
uncertainties, the $J/\psi$ $A_N$ is consistent with zero in all $p_T$ 
and $x_F$ bins in $p$$+$Al collisions. For $p$$+$Au 
collisions, the data favor negative asymmetries in all $x_F$ bins and we 
have observed a nonzero $A_N$ at the 2$\sigma$ level in the lower-$p_T$ 
bins; however, in the higher-$p_T$ bins, it is consistent with zero. 
This intriguing result observed in $p$$+$Au collisions could 
indicate possible contributions from other nonconventional mechanisms. 
One of the possible contributions could come from electromagnetic 
interactions. A recent PHENIX measurement~\cite{Aidala:2017cnz} of the 
TSSA in forward neutron production shows that electromagnetic processes 
could significantly enhance $A_N$ in $p$+$A$ collisions, resulting in a 
strong nuclear-size dependence for $A_N$. Further theoretical studies of 
$A_N$ in $J/\psi$ production exploring different mechanisms are needed 
to explain the current results.

\section*{ACKNOWLEDGMENTS}   

We thank the staff of the Collider-Accelerator and Physics
Departments at Brookhaven National Laboratory and the staff of
the other PHENIX participating institutions for their vital
contributions.  We acknowledge support from the
Medium Energy Nuclear Physics Program in the 
Office of Nuclear Physics in the
Office of Science of the Department of Energy,
the National Science Foundation,
Abilene Christian University Research Council,
Research Foundation of SUNY, and
Dean of the College of Arts and Sciences, Vanderbilt University
(U.S.A),
Ministry of Education, Culture, Sports, Science, and Technology
and the Japan Society for the Promotion of Science (Japan),
Conselho Nacional de Desenvolvimento Cient\'{\i}fico e
Tecnol{\'o}gico and Funda\c c{\~a}o de Amparo {\`a} Pesquisa do
Estado de S{\~a}o Paulo (Brazil),
Natural Science Foundation of China (People's Republic of China),
Croatian Science Foundation and
Ministry of Science and Education (Croatia),
Ministry of Education, Youth and Sports (Czech Republic),
Centre National de la Recherche Scientifique, Commissariat
{\`a} l'{\'E}nergie Atomique, and Institut National de Physique
Nucl{\'e}aire et de Physique des Particules (France),
Bundesministerium f\"ur Bildung und Forschung, Deutscher
Akademischer Austausch Dienst, and Alexander von Humboldt Stiftung (Germany),
J. Bolyai Research Scholarship, EFOP, the New National Excellence
Program ({\'U}NKP), NKFIH, and OTKA (Hungary),
Department of Atomic Energy and Department of Science and Technology (India),
Israel Science Foundation (Israel),
Basic Science Research Program through NRF of the Ministry of Education (Korea),
Physics Department, Lahore University of Management Sciences (Pakistan),
Ministry of Education and Science, Russian Academy of Sciences,
Federal Agency of Atomic Energy (Russia),
VR and Wallenberg Foundation (Sweden),
the U.S. Civilian Research and Development Foundation for the
Independent States of the Former Soviet Union,
the Hungarian American Enterprise Scholarship Fund,
the US-Hungarian Fulbright Foundation,
and the US-Israel Binational Science Foundation.




\begin{thebibliography}{62}%
\makeatletter
\providecommand \@ifxundefined [1]{%
 \@ifx{#1\undefined}
}%
\providecommand \@ifnum [1]{%
 \ifnum #1\expandafter \@firstoftwo
 \else \expandafter \@secondoftwo
 \fi
}%
\providecommand \@ifx [1]{%
 \ifx #1\expandafter \@firstoftwo
 \else \expandafter \@secondoftwo
 \fi
}%
\providecommand \natexlab [1]{#1}%
\providecommand \enquote  [1]{``#1''}%
\providecommand \bibnamefont  [1]{#1}%
\providecommand \bibfnamefont [1]{#1}%
\providecommand \citenamefont [1]{#1}%
\providecommand \href@noop [0]{\@secondoftwo}%
\providecommand \href [0]{\begingroup \@sanitize@url \@href}%
\providecommand \@href[1]{\@@startlink{#1}\@@href}%
\providecommand \@@href[1]{\endgroup#1\@@endlink}%
\providecommand \@sanitize@url [0]{\catcode `\\12\catcode `\$12\catcode
  `\&12\catcode `\#12\catcode `\^12\catcode `\_12\catcode `\%12\relax}%
\providecommand \@@startlink[1]{}%
\providecommand \@@endlink[0]{}%
\providecommand \url  [0]{\begingroup\@sanitize@url \@url }%
\providecommand \@url [1]{\endgroup\@href {#1}{\urlprefix }}%
\providecommand \urlprefix  [0]{URL }%
\providecommand \Eprint [0]{\href }%
\providecommand \doibase [0]{http://dx.doi.org/}%
\providecommand \selectlanguage [0]{\@gobble}%
\providecommand \bibinfo  [0]{\@secondoftwo}%
\providecommand \bibfield  [0]{\@secondoftwo}%
\providecommand \translation [1]{[#1]}%
\providecommand \BibitemOpen [0]{}%
\providecommand \bibitemStop [0]{}%
\providecommand \bibitemNoStop [0]{.\EOS\space}%
\providecommand \EOS [0]{\spacefactor3000\relax}%
\providecommand \BibitemShut  [1]{\csname bibitem#1\endcsname}%
\let\auto@bib@innerbib\@empty
\bibitem [{\citenamefont {Kane}\ \emph {et~al.}(1978)\citenamefont {Kane},
  \citenamefont {Pumplin},\ and\ \citenamefont {Repko}}]{Kane:1978nd}%
  \BibitemOpen
  \bibfield  {author} {\bibinfo {author} {\bibfnamefont {G.~L.}\ \bibnamefont
  {Kane}}, \bibinfo {author} {\bibfnamefont {J.}~\bibnamefont {Pumplin}}, \
  and\ \bibinfo {author} {\bibfnamefont {W.}~\bibnamefont {Repko}},\ }\bibfield
   {title} {\enquote {\bibinfo {title} {{Transverse Quark Polarization in Large
  p(T) Reactions, e+ e- Jets, and Leptoproduction: A Test of QCD}},}\ }\href
  {\doibase 10.1103/PhysRevLett.41.1689} {\bibfield  {journal} {\bibinfo
  {journal} {Phys. Rev. Lett.}\ }\textbf {\bibinfo {volume} {41}},\ \bibinfo
  {pages} {1689} (\bibinfo {year} {1978})}\BibitemShut {NoStop}%
\bibitem [{\citenamefont {Klem}\ \emph {et~al.}(1976)\citenamefont {Klem},
  \citenamefont {Bowers}, \citenamefont {Courant}, \citenamefont {Kagan},
  \citenamefont {Marshak}, \citenamefont {Peterson}, \citenamefont {Ruddick},
  \citenamefont {Dragoset},\ and\ \citenamefont {Roberts}}]{Klem:1976ui}%
  \BibitemOpen
  \bibfield  {author} {\bibinfo {author} {\bibfnamefont {R.~D.}\ \bibnamefont
  {Klem}}, \bibinfo {author} {\bibfnamefont {J.~E.}\ \bibnamefont {Bowers}},
  \bibinfo {author} {\bibfnamefont {H.~W.}\ \bibnamefont {Courant}}, \bibinfo
  {author} {\bibfnamefont {H.}~\bibnamefont {Kagan}}, \bibinfo {author}
  {\bibfnamefont {M.~L.}\ \bibnamefont {Marshak}}, \bibinfo {author}
  {\bibfnamefont {E.~A.}\ \bibnamefont {Peterson}}, \bibinfo {author}
  {\bibfnamefont {K.}~\bibnamefont {Ruddick}}, \bibinfo {author} {\bibfnamefont
  {W.~H.}\ \bibnamefont {Dragoset}}, \ and\ \bibinfo {author} {\bibfnamefont
  {J.~B.}\ \bibnamefont {Roberts}},\ }\bibfield  {title} {\enquote {\bibinfo
  {title} {{Measurement of Asymmetries of Inclusive Pion Production in Proton
  Proton Interactions at 6-GeV/c and 11.8-GeV/c}},}\ }\href {\doibase
  10.1103/PhysRevLett.36.929} {\bibfield  {journal} {\bibinfo  {journal} {Phys.
  Rev. Lett.}\ }\textbf {\bibinfo {volume} {36}},\ \bibinfo {pages} {929--931}
  (\bibinfo {year} {1976})}\BibitemShut {NoStop}%
\bibitem [{\citenamefont {Aschenauer}\ \emph {et~al.}()\citenamefont
  {Aschenauer} \emph {et~al.}}]{Aschenauer:2015eha}%
  \BibitemOpen
  \bibfield  {author} {\bibinfo {author} {\bibfnamefont {E.-C.}\ \bibnamefont
  {Aschenauer}} \emph {et~al.},\ }\href@noop {} {\enquote {\bibinfo {title}
  {{The RHIC SPIN Program: Achievements and Future Opportunities}},}\ }\bibinfo
  {note} {ArXiv:1501.01220}\BibitemShut {NoStop}%
\bibitem [{\citenamefont {Sivers}(1990)}]{Sivers:1989cc}%
  \BibitemOpen
  \bibfield  {author} {\bibinfo {author} {\bibfnamefont {D.~W.}\ \bibnamefont
  {Sivers}},\ }\bibfield  {title} {\enquote {\bibinfo {title} {{Single Spin
  Production Asymmetries from the Hard Scattering of Point-Like
  Constituents}},}\ }\href {\doibase 10.1103/PhysRevD.41.83} {\bibfield
  {journal} {\bibinfo  {journal} {Phys. Rev. D}\ }\textbf {\bibinfo {volume}
  {41}},\ \bibinfo {pages} {83} (\bibinfo {year} {1990})}\BibitemShut {NoStop}%
\bibitem [{\citenamefont {Sivers}(1991)}]{Sivers:1990fh}%
  \BibitemOpen
  \bibfield  {author} {\bibinfo {author} {\bibfnamefont {D.~W.}\ \bibnamefont
  {Sivers}},\ }\bibfield  {title} {\enquote {\bibinfo {title} {{Hard scattering
  scaling laws for single spin production asymmetries}},}\ }\href {\doibase
  10.1103/PhysRevD.43.261} {\bibfield  {journal} {\bibinfo  {journal} {Phys.
  Rev. D}\ }\textbf {\bibinfo {volume} {43}},\ \bibinfo {pages} {261--263}
  (\bibinfo {year} {1991})}\BibitemShut {NoStop}%
\bibitem [{\citenamefont {Collins}(1993)}]{Collins:1992kk}%
  \BibitemOpen
  \bibfield  {author} {\bibinfo {author} {\bibfnamefont {J.~C.}\ \bibnamefont
  {Collins}},\ }\bibfield  {title} {\enquote {\bibinfo {title} {{Fragmentation
  of transversely polarized quarks probed in transverse momentum
  distributions}},}\ }\href {\doibase 10.1016/0550-3213(93)90262-N} {\bibfield
  {journal} {\bibinfo  {journal} {Nucl. Phys. B}\ }\textbf {\bibinfo {volume}
  {396}},\ \bibinfo {pages} {161} (\bibinfo {year} {1993})}\BibitemShut
  {NoStop}%
\bibitem [{\citenamefont {Efremov}\ and\ \citenamefont
  {Teryaev}(1985)}]{Efremov:1984ip}%
  \BibitemOpen
  \bibfield  {author} {\bibinfo {author} {\bibfnamefont {A.~V.}\ \bibnamefont
  {Efremov}}\ and\ \bibinfo {author} {\bibfnamefont {O.~V.}\ \bibnamefont
  {Teryaev}},\ }\bibfield  {title} {\enquote {\bibinfo {title} {{QCD Asymmetry
  and Polarized Hadron Structure Functions}},}\ }\bibfield  {booktitle} {\emph
  {\bibinfo {booktitle} {{Workshop on High-Energy Spin Physics Protvino, USSR,
  September 14-17, 1983}}},\ }\href {\doibase 10.1016/0370-2693(85)90999-2}
  {\bibfield  {journal} {\bibinfo  {journal} {Phys. Lett. B}\ }\textbf
  {\bibinfo {volume} {150}},\ \bibinfo {pages} {383} (\bibinfo {year}
  {1985})}\BibitemShut {NoStop}%
\bibitem [{\citenamefont {Qiu}\ and\ \citenamefont
  {Sterman}(1991)}]{Qiu:1991pp}%
  \BibitemOpen
  \bibfield  {author} {\bibinfo {author} {\bibfnamefont {J.-W.}\ \bibnamefont
  {Qiu}}\ and\ \bibinfo {author} {\bibfnamefont {G.~F.}\ \bibnamefont
  {Sterman}},\ }\bibfield  {title} {\enquote {\bibinfo {title} {{Single
  transverse spin asymmetries}},}\ }\href {\doibase
  10.1103/PhysRevLett.67.2264} {\bibfield  {journal} {\bibinfo  {journal}
  {Phys. Rev. Lett.}\ }\textbf {\bibinfo {volume} {67}},\ \bibinfo {pages}
  {2264} (\bibinfo {year} {1991})}\BibitemShut {NoStop}%
\bibitem [{\citenamefont {Qiu}\ and\ \citenamefont
  {Sterman}(1992)}]{Qiu:1991wg}%
  \BibitemOpen
  \bibfield  {author} {\bibinfo {author} {\bibfnamefont {J.-W.}\ \bibnamefont
  {Qiu}}\ and\ \bibinfo {author} {\bibfnamefont {G.~F.}\ \bibnamefont
  {Sterman}},\ }\bibfield  {title} {\enquote {\bibinfo {title} {{Single
  transverse spin asymmetries in direct photon production}},}\ }\href {\doibase
  10.1016/0550-3213(92)90003-T} {\bibfield  {journal} {\bibinfo  {journal}
  {Nucl. Phys B.}\ }\textbf {\bibinfo {volume} {378}},\ \bibinfo {pages} {52}
  (\bibinfo {year} {1992})}\BibitemShut {NoStop}%
\bibitem [{\citenamefont {Kang}\ \emph
  {et~al.}(2010{\natexlab{a}})\citenamefont {Kang}, \citenamefont {Yuan},\ and\
  \citenamefont {Zhou}}]{Kang:2010zzb}%
  \BibitemOpen
  \bibfield  {author} {\bibinfo {author} {\bibfnamefont {Z.-B.}\ \bibnamefont
  {Kang}}, \bibinfo {author} {\bibfnamefont {F.}~\bibnamefont {Yuan}}, \ and\
  \bibinfo {author} {\bibfnamefont {J.}~\bibnamefont {Zhou}},\ }\bibfield
  {title} {\enquote {\bibinfo {title} {{Twist-three fragmentation function
  contribution to the single spin asymmetry in p p collisions}},}\ }\href
  {\doibase 10.1016/j.physletb.2010.07.003} {\bibfield  {journal} {\bibinfo
  {journal} {Phys. Lett. B}\ }\textbf {\bibinfo {volume} {691}},\ \bibinfo
  {pages} {243} (\bibinfo {year} {2010}{\natexlab{a}})}\BibitemShut {NoStop}%
\bibitem [{\citenamefont {Kang}\ and\ \citenamefont {Qiu}(2009)}]{Kang:2008ey}%
  \BibitemOpen
  \bibfield  {author} {\bibinfo {author} {\bibfnamefont {Z.-B.}\ \bibnamefont
  {Kang}}\ and\ \bibinfo {author} {\bibfnamefont {J.-W.}\ \bibnamefont {Qiu}},\
  }\bibfield  {title} {\enquote {\bibinfo {title} {{Evolution of twist-3
  multi-parton correlation functions relevant to single transverse-spin
  asymmetry}},}\ }\href {\doibase 10.1103/PhysRevD.79.016003} {\bibfield
  {journal} {\bibinfo  {journal} {Phys. Rev. D}\ }\textbf {\bibinfo {volume}
  {79}},\ \bibinfo {pages} {016003} (\bibinfo {year} {2009})}\BibitemShut
  {NoStop}%
\bibitem [{\citenamefont {Kang}\ and\ \citenamefont {Qiu}(2008)}]{Kang:2008qh}%
  \BibitemOpen
  \bibfield  {author} {\bibinfo {author} {\bibfnamefont {Z.-B.}\ \bibnamefont
  {Kang}}\ and\ \bibinfo {author} {\bibfnamefont {J.-W.}\ \bibnamefont {Qiu}},\
  }\bibfield  {title} {\enquote {\bibinfo {title} {{Single transverse-spin
  asymmetry for D-meson production in semi-inclusive deep inelastic
  scattering}},}\ }\href {\doibase 10.1103/PhysRevD.78.034005} {\bibfield
  {journal} {\bibinfo  {journal} {Phys. Rev. D}\ }\textbf {\bibinfo {volume}
  {78}},\ \bibinfo {pages} {034005} (\bibinfo {year} {2008})}\BibitemShut
  {NoStop}%
\bibitem [{\citenamefont {Qiu}\ and\ \citenamefont
  {Sterman}(1998)}]{PhysRevD.59.014004}%
  \BibitemOpen
  \bibfield  {author} {\bibinfo {author} {\bibfnamefont {J.}~\bibnamefont
  {Qiu}}\ and\ \bibinfo {author} {\bibfnamefont {G.}~\bibnamefont {Sterman}},\
  }\bibfield  {title} {\enquote {\bibinfo {title} {Single transverse-spin
  asymmetries in hadronic pion production},}\ }\href {\doibase
  10.1103/PhysRevD.59.014004} {\bibfield  {journal} {\bibinfo  {journal} {Phys.
  Rev. D}\ }\textbf {\bibinfo {volume} {59}},\ \bibinfo {pages} {014004}
  (\bibinfo {year} {1998})}\BibitemShut {NoStop}%
\bibitem [{\citenamefont {Kang}\ \emph
  {et~al.}(2010{\natexlab{b}})\citenamefont {Kang}, \citenamefont {Qiu},\ and\
  \citenamefont {Zhang}}]{Kang:2010hg}%
  \BibitemOpen
  \bibfield  {author} {\bibinfo {author} {\bibfnamefont {Z.-B.}\ \bibnamefont
  {Kang}}, \bibinfo {author} {\bibfnamefont {J.-W.}\ \bibnamefont {Qiu}}, \
  and\ \bibinfo {author} {\bibfnamefont {H.}~\bibnamefont {Zhang}},\ }\bibfield
   {title} {\enquote {\bibinfo {title} {{Quark-gluon correlation functions
  relevant to single transverse spin asymmetries}},}\ }\href {\doibase
  10.1103/PhysRevD.81.114030} {\bibfield  {journal} {\bibinfo  {journal} {Phys.
  Rev. D}\ }\textbf {\bibinfo {volume} {81}},\ \bibinfo {pages} {114030}
  (\bibinfo {year} {2010}{\natexlab{b}})}\BibitemShut {NoStop}%
\bibitem [{\citenamefont {Braun}\ \emph {et~al.}(2009)\citenamefont {Braun},
  \citenamefont {Manashov},\ and\ \citenamefont {Pirnay}}]{Braun:2009mi}%
  \BibitemOpen
  \bibfield  {author} {\bibinfo {author} {\bibfnamefont {V.~M.}\ \bibnamefont
  {Braun}}, \bibinfo {author} {\bibfnamefont {A.~N.}\ \bibnamefont {Manashov}},
  \ and\ \bibinfo {author} {\bibfnamefont {B.}~\bibnamefont {Pirnay}},\
  }\bibfield  {title} {\enquote {\bibinfo {title} {{Scale dependence of
  twist-three contributions to single spin asymmetries}},}\ }\href {\doibase
  10.1103/PhysRevD.80.114002, 10.1103/PhysRevD.86.119902} {\bibfield  {journal}
  {\bibinfo  {journal} {Phys. Rev. D}\ }\textbf {\bibinfo {volume} {80}},\
  \bibinfo {pages} {114002} (\bibinfo {year} {2009})},\ \bibinfo {note}
  {[Erratum: Phys. Rev. D {\bf 86}, 119902(E) (2012)]}\BibitemShut {NoStop}%
\bibitem [{\citenamefont {Adams}\ \emph {et~al.}(2004)\citenamefont {Adams}
  \emph {et~al.}}]{Adams:2003fx}%
  \BibitemOpen
  \bibfield  {author} {\bibinfo {author} {\bibfnamefont {J.}~\bibnamefont
  {Adams}} \emph {et~al.} (\bibinfo {collaboration} {STAR Collaboration}),\
  }\bibfield  {title} {\enquote {\bibinfo {title} {{Cross-sections and
  transverse single spin asymmetries in forward neutral pion production from
  proton collisions at $\sqrt{s}=200$ GeV}},}\ }\href {\doibase
  10.1103/PhysRevLett.92.171801} {\bibfield  {journal} {\bibinfo  {journal}
  {Phys. Rev. Lett.}\ }\textbf {\bibinfo {volume} {92}},\ \bibinfo {pages}
  {171801} (\bibinfo {year} {2004})}\BibitemShut {NoStop}%
\bibitem [{\citenamefont {Abelev}\ \emph {et~al.}(2008)\citenamefont {Abelev}
  \emph {et~al.}}]{Abelev:2008af}%
  \BibitemOpen
  \bibfield  {author} {\bibinfo {author} {\bibfnamefont {B.~I.}\ \bibnamefont
  {Abelev}} \emph {et~al.} (\bibinfo {collaboration} {STAR Collaboration}),\
  }\bibfield  {title} {\enquote {\bibinfo {title} {{Forward Neutral Pion
  Transverse Single Spin Asymmetries in p+p Collisions at $\sqrt{s}=200$
  GeV}},}\ }\href {\doibase 10.1103/PhysRevLett.101.222001} {\bibfield
  {journal} {\bibinfo  {journal} {Phys. Rev. Lett.}\ }\textbf {\bibinfo
  {volume} {101}},\ \bibinfo {pages} {222001} (\bibinfo {year}
  {2008})}\BibitemShut {NoStop}%
\bibitem [{\citenamefont {Adamczyk}\ \emph {et~al.}(2012)\citenamefont
  {Adamczyk} \emph {et~al.}}]{Adamczyk:2012xd}%
  \BibitemOpen
  \bibfield  {author} {\bibinfo {author} {\bibfnamefont {L.}~\bibnamefont
  {Adamczyk}} \emph {et~al.} (\bibinfo {collaboration} {STAR Collaboration}),\
  }\bibfield  {title} {\enquote {\bibinfo {title} {{Transverse Single-Spin
  Asymmetry and Cross-Section for $\pi^0$ and $\eta$ Mesons at Large
  Feynman-$x$ in Polarized $p+p$ Collisions at $\sqrt{s}=200$ GeV}},}\ }\href
  {\doibase 10.1103/PhysRevD.86.051101} {\bibfield  {journal} {\bibinfo
  {journal} {Phys. Rev. D}\ }\textbf {\bibinfo {volume} {86}},\ \bibinfo
  {pages} {051101} (\bibinfo {year} {2012})}\BibitemShut {NoStop}%
\bibitem [{\citenamefont {Adler}\ \emph {et~al.}(2005)\citenamefont {Adler}
  \emph {et~al.}}]{Adler:2005in}%
  \BibitemOpen
  \bibfield  {author} {\bibinfo {author} {\bibfnamefont {S.~S.}\ \bibnamefont
  {Adler}} \emph {et~al.} (\bibinfo {collaboration} {PHENIX Collaboration}),\
  }\bibfield  {title} {\enquote {\bibinfo {title} {{Measurement of transverse
  single-spin asymmetries for mid-rapidity production of neutral pions and
  charged hadrons in polarized p+p collisions at $\sqrt{s}=200$ GeV}},}\ }\href
  {\doibase 10.1103/PhysRevLett.95.202001} {\bibfield  {journal} {\bibinfo
  {journal} {Phys. Rev. Lett.}\ }\textbf {\bibinfo {volume} {95}},\ \bibinfo
  {pages} {202001} (\bibinfo {year} {2005})}\BibitemShut {NoStop}%
\bibitem [{\citenamefont {Adare}\ \emph
  {et~al.}(2014{\natexlab{a}})\citenamefont {Adare} \emph
  {et~al.}}]{Adare:2014qzo}%
  \BibitemOpen
  \bibfield  {author} {\bibinfo {author} {\bibfnamefont {A.}~\bibnamefont
  {Adare}} \emph {et~al.} (\bibinfo {collaboration} {PHENIX Collaboration}),\
  }\bibfield  {title} {\enquote {\bibinfo {title} {{Cross section and
  transverse single-spin asymmetry of $\eta$ mesons in $p^{\uparrow}+p$
  collisions at $\sqrt{s}=200$ GeV at forward rapidity}},}\ }\href {\doibase
  10.1103/PhysRevD.90.072008} {\bibfield  {journal} {\bibinfo  {journal} {Phys.
  Rev. D}\ }\textbf {\bibinfo {volume} {90}},\ \bibinfo {pages} {072008}
  (\bibinfo {year} {2014}{\natexlab{a}})}\BibitemShut {NoStop}%
\bibitem [{\citenamefont {Adare}\ \emph
  {et~al.}(2014{\natexlab{b}})\citenamefont {Adare} \emph
  {et~al.}}]{Adare:2013ekj}%
  \BibitemOpen
  \bibfield  {author} {\bibinfo {author} {\bibfnamefont {A.}~\bibnamefont
  {Adare}} \emph {et~al.} (\bibinfo {collaboration} {PHENIX Collaboration}),\
  }\bibfield  {title} {\enquote {\bibinfo {title} {{Measurement of
  transverse-single-spin asymmetries for midrapidity and forward-rapidity
  production of hadrons in polarized p+p collisions at $\sqrt{s}=200$ and 62.4
  GeV}},}\ }\href {\doibase 10.1103/PhysRevD.90.012006} {\bibfield  {journal}
  {\bibinfo  {journal} {Phys. Rev. D}\ }\textbf {\bibinfo {volume} {90}},\
  \bibinfo {pages} {012006} (\bibinfo {year} {2014}{\natexlab{b}})}\BibitemShut
  {NoStop}%
\bibitem [{\citenamefont {Aidala}\ \emph {et~al.}(2017)\citenamefont {Aidala}
  \emph {et~al.}}]{Aidala:2017pum}%
  \BibitemOpen
  \bibfield  {author} {\bibinfo {author} {\bibfnamefont {C.}~\bibnamefont
  {Aidala}} \emph {et~al.} (\bibinfo {collaboration} {PHENIX Collaboration}),\
  }\bibfield  {title} {\enquote {\bibinfo {title} {{Cross section and
  transverse single-spin asymmetry of muons from open heavy-flavor decays in
  polarized $p$+$p$ collisions at $\sqrt{s}=200$ GeV}},}\ }\href {\doibase
  10.1103/PhysRevD.95.112001} {\bibfield  {journal} {\bibinfo  {journal} {Phys.
  Rev. D}\ }\textbf {\bibinfo {volume} {95}},\ \bibinfo {pages} {112001}
  (\bibinfo {year} {2017})}\BibitemShut {NoStop}%
\bibitem [{\citenamefont {Arsene}\ \emph {et~al.}(2008)\citenamefont {Arsene}
  \emph {et~al.}}]{Arsene:2008aa}%
  \BibitemOpen
  \bibfield  {author} {\bibinfo {author} {\bibfnamefont {I.}~\bibnamefont
  {Arsene}} \emph {et~al.} (\bibinfo {collaboration} {BRAHMS Collaboration}),\
  }\bibfield  {title} {\enquote {\bibinfo {title} {{Single Transverse Spin
  Asymmetries of Identified Charged Hadrons in Polarized p+p Collisions at
  $\sqrt{s}=62.4$ GeV}},}\ }\href {\doibase 10.1103/PhysRevLett.101.042001}
  {\bibfield  {journal} {\bibinfo  {journal} {Phys. Rev. Lett.}\ }\textbf
  {\bibinfo {volume} {101}},\ \bibinfo {pages} {042001} (\bibinfo {year}
  {2008})}\BibitemShut {NoStop}%
\bibitem [{\citenamefont {Anselmino}\ \emph {et~al.}(2012)\citenamefont
  {Anselmino}, \citenamefont {Boglione}, \citenamefont {D'Alesio},
  \citenamefont {Leader}, \citenamefont {Melis}, \citenamefont {Murgia},\ and\
  \citenamefont {Prokudin}}]{Anselmino:2012rq}%
  \BibitemOpen
  \bibfield  {author} {\bibinfo {author} {\bibfnamefont {M.}~\bibnamefont
  {Anselmino}}, \bibinfo {author} {\bibfnamefont {M.}~\bibnamefont {Boglione}},
  \bibinfo {author} {\bibfnamefont {U.}~\bibnamefont {D'Alesio}}, \bibinfo
  {author} {\bibfnamefont {E.}~\bibnamefont {Leader}}, \bibinfo {author}
  {\bibfnamefont {S.}~\bibnamefont {Melis}}, \bibinfo {author} {\bibfnamefont
  {F.}~\bibnamefont {Murgia}}, \ and\ \bibinfo {author} {\bibfnamefont
  {A.}~\bibnamefont {Prokudin}},\ }\bibfield  {title} {\enquote {\bibinfo
  {title} {{On the role of Collins effect in the single spin asymmetry $A_N$ in
  $p^\uparrow p \to h X$ processes}},}\ }\href {\doibase
  10.1103/PhysRevD.86.074032} {\bibfield  {journal} {\bibinfo  {journal} {Phys.
  Rev. D}\ }\textbf {\bibinfo {volume} {86}},\ \bibinfo {pages} {074032}
  (\bibinfo {year} {2012})}\BibitemShut {NoStop}%
\bibitem [{\citenamefont {Collins}\ \emph {et~al.}(1985)\citenamefont
  {Collins}, \citenamefont {Soper},\ and\ \citenamefont
  {Sterman}}]{Collins:1984kg}%
  \BibitemOpen
  \bibfield  {author} {\bibinfo {author} {\bibfnamefont {J.~C.}\ \bibnamefont
  {Collins}}, \bibinfo {author} {\bibfnamefont {D.~E.}\ \bibnamefont {Soper}},
  \ and\ \bibinfo {author} {\bibfnamefont {G.~F.}\ \bibnamefont {Sterman}},\
  }\bibfield  {title} {\enquote {\bibinfo {title} {{Transverse Momentum
  Distribution in Drell-Yan Pair and W and Z Boson Production}},}\ }\href
  {\doibase 10.1016/0550-3213(85)90479-1} {\bibfield  {journal} {\bibinfo
  {journal} {Nucl. Phys. B}\ }\textbf {\bibinfo {volume} {250}},\ \bibinfo
  {pages} {199} (\bibinfo {year} {1985})}\BibitemShut {NoStop}%
\bibitem [{\citenamefont {Airapetian}\ \emph {et~al.}(2005)\citenamefont
  {Airapetian} \emph {et~al.}}]{Airapetian:2004tw}%
  \BibitemOpen
  \bibfield  {author} {\bibinfo {author} {\bibfnamefont {A.}~\bibnamefont
  {Airapetian}} \emph {et~al.} (\bibinfo {collaboration} {HERMES
  Collaboration}),\ }\bibfield  {title} {\enquote {\bibinfo {title}
  {{Single-spin asymmetries in semi-inclusive deep-inelastic scattering on a
  transversely polarized hydrogen target}},}\ }\href {\doibase
  10.1103/PhysRevLett.94.012002} {\bibfield  {journal} {\bibinfo  {journal}
  {Phys. Rev. Lett.}\ }\textbf {\bibinfo {volume} {94}},\ \bibinfo {pages}
  {012002} (\bibinfo {year} {2005})}\BibitemShut {NoStop}%
\bibitem [{\citenamefont {Adolph}\ \emph {et~al.}(2017)\citenamefont {Adolph}
  \emph {et~al.}}]{Adolph:2017pgv}%
  \BibitemOpen
  \bibfield  {author} {\bibinfo {author} {\bibfnamefont {C.}~\bibnamefont
  {Adolph}} \emph {et~al.} (\bibinfo {collaboration} {COMPASS Collaboration}),\
  }\bibfield  {title} {\enquote {\bibinfo {title} {{First measurement of the
  Sivers asymmetry for gluons using SIDIS data}},}\ }\href {\doibase
  10.1016/j.physletb.2017.07.018} {\bibfield  {journal} {\bibinfo  {journal}
  {Phys. Lett. B}\ }\textbf {\bibinfo {volume} {772}},\ \bibinfo {pages} {854}
  (\bibinfo {year} {2017})}\BibitemShut {NoStop}%
\bibitem [{\citenamefont {Godbole}\ \emph
  {et~al.}(2017{\natexlab{a}})\citenamefont {Godbole}, \citenamefont {Kaushik},
  \citenamefont {Misra}, \citenamefont {Rawoot},\ and\ \citenamefont
  {Sonawane}}]{Godbole:2016ixc}%
  \BibitemOpen
  \bibfield  {author} {\bibinfo {author} {\bibfnamefont {R.~M.}\ \bibnamefont
  {Godbole}}, \bibinfo {author} {\bibfnamefont {A.}~\bibnamefont {Kaushik}},
  \bibinfo {author} {\bibfnamefont {A.}~\bibnamefont {Misra}}, \bibinfo
  {author} {\bibfnamefont {V.}~\bibnamefont {Rawoot}}, \ and\ \bibinfo {author}
  {\bibfnamefont {B.}~\bibnamefont {Sonawane}},\ }\bibfield  {title} {\enquote
  {\bibinfo {title} {{Heavy Flavour production as probe of Gluon Sivers
  Function}},}\ }\bibfield  {booktitle} {\emph {\bibinfo {booktitle}
  {{Proceedings, Theory and Experiment for Hadrons on the Light-Front (Light
  Cone 2016): Lisbon, Portugal, September 5-8, 2016}}},\ }\href {\doibase
  10.1007/s00601-017-1256-8} {\bibfield  {journal} {\bibinfo  {journal} {Few
  Body Syst.}\ }\textbf {\bibinfo {volume} {58}},\ \bibinfo {pages} {96}
  (\bibinfo {year} {2017}{\natexlab{a}})}\BibitemShut {NoStop}%
\bibitem [{\citenamefont {D'Alesio}\ \emph {et~al.}()\citenamefont {D'Alesio},
  \citenamefont {Murgia},\ and\ \citenamefont {Pisano}}]{DAlesio:2015fwo}%
  \BibitemOpen
  \bibfield  {author} {\bibinfo {author} {\bibfnamefont {U.}~\bibnamefont
  {D'Alesio}}, \bibinfo {author} {\bibfnamefont {F.}~\bibnamefont {Murgia}}, \
  and\ \bibinfo {author} {\bibfnamefont {C.}~\bibnamefont {Pisano}},\
  }\href@noop {} {\enquote {\bibinfo {title} {{Towards a first estimate of the
  gluon Sivers function from A$_{N}$ data in pp collisions at RHIC}},}\
  }\bibinfo {note} {{J. High Energy Phys. {\bf 09 (2015)} 119}}\BibitemShut
  {NoStop}%
\bibitem [{\citenamefont {Mukherjee}\ and\ \citenamefont
  {Rajesh}(2017)}]{Mukherjee:2016qxa}%
  \BibitemOpen
  \bibfield  {author} {\bibinfo {author} {\bibfnamefont {A.}~\bibnamefont
  {Mukherjee}}\ and\ \bibinfo {author} {\bibfnamefont {S.}~\bibnamefont
  {Rajesh}},\ }\bibfield  {title} {\enquote {\bibinfo {title} {{$J/\psi $
  production in polarized and unpolarized ep collision and Sivers and $\cos
  2\phi $ asymmetries}},}\ }\href {\doibase 10.1140/epjc/s10052-017-5406-4}
  {\bibfield  {journal} {\bibinfo  {journal} {Eur. Phys. J. C}\ }\textbf
  {\bibinfo {volume} {77}},\ \bibinfo {pages} {854} (\bibinfo {year}
  {2017})}\BibitemShut {NoStop}%
\bibitem [{\citenamefont {Boer}\ \emph {et~al.}(2003)\citenamefont {Boer},
  \citenamefont {Mulders},\ and\ \citenamefont {Pijlman}}]{Boer:2003cm}%
  \BibitemOpen
  \bibfield  {author} {\bibinfo {author} {\bibfnamefont {D.}~\bibnamefont
  {Boer}}, \bibinfo {author} {\bibfnamefont {P.~J.}\ \bibnamefont {Mulders}}, \
  and\ \bibinfo {author} {\bibfnamefont {F.}~\bibnamefont {Pijlman}},\
  }\bibfield  {title} {\enquote {\bibinfo {title} {{Universality of T odd
  effects in single spin and azimuthal asymmetries}},}\ }\href {\doibase
  10.1016/S0550-3213(03)00527-3} {\bibfield  {journal} {\bibinfo  {journal}
  {Nucl. Phys. B}\ }\textbf {\bibinfo {volume} {667}},\ \bibinfo {pages} {201}
  (\bibinfo {year} {2003})}\BibitemShut {NoStop}%
\bibitem [{\citenamefont {Anselmino}\ \emph {et~al.}(2013)\citenamefont
  {Anselmino}, \citenamefont {Boglione}, \citenamefont {D'Alesio},
  \citenamefont {Melis}, \citenamefont {Murgia},\ and\ \citenamefont
  {Prokudin}}]{Anselmino:2013rya}%
  \BibitemOpen
  \bibfield  {author} {\bibinfo {author} {\bibfnamefont {M.}~\bibnamefont
  {Anselmino}}, \bibinfo {author} {\bibfnamefont {M.}~\bibnamefont {Boglione}},
  \bibinfo {author} {\bibfnamefont {U.}~\bibnamefont {D'Alesio}}, \bibinfo
  {author} {\bibfnamefont {S.}~\bibnamefont {Melis}}, \bibinfo {author}
  {\bibfnamefont {F.}~\bibnamefont {Murgia}}, \ and\ \bibinfo {author}
  {\bibfnamefont {A.}~\bibnamefont {Prokudin}},\ }\bibfield  {title} {\enquote
  {\bibinfo {title} {{Sivers effect and the single spin asymmetry $A_{N}$ in
  $p^{\uparrow} p \to hX$ processes}},}\ }\href {\doibase
  10.1103/PhysRevD.88.054023} {\bibfield  {journal} {\bibinfo  {journal} {Phys.
  Rev. D}\ }\textbf {\bibinfo {volume} {88}},\ \bibinfo {pages} {054023}
  (\bibinfo {year} {2013})}\BibitemShut {NoStop}%
\bibitem [{\citenamefont {Anselmino}\ \emph {et~al.}(2009)\citenamefont
  {Anselmino}, \citenamefont {Boglione}, \citenamefont {D'Alesio},
  \citenamefont {Kotzinian}, \citenamefont {Melis}, \citenamefont {Murgia},
  \citenamefont {Prokudin},\ and\ \citenamefont {Turk}}]{Anselmino:2008sga}%
  \BibitemOpen
  \bibfield  {author} {\bibinfo {author} {\bibfnamefont {M.}~\bibnamefont
  {Anselmino}}, \bibinfo {author} {\bibfnamefont {M.}~\bibnamefont {Boglione}},
  \bibinfo {author} {\bibfnamefont {U.}~\bibnamefont {D'Alesio}}, \bibinfo
  {author} {\bibfnamefont {A.}~\bibnamefont {Kotzinian}}, \bibinfo {author}
  {\bibfnamefont {S.}~\bibnamefont {Melis}}, \bibinfo {author} {\bibfnamefont
  {F.}~\bibnamefont {Murgia}}, \bibinfo {author} {\bibfnamefont
  {A.}~\bibnamefont {Prokudin}}, \ and\ \bibinfo {author} {\bibfnamefont
  {C.}~\bibnamefont {Turk}},\ }\bibfield  {title} {\enquote {\bibinfo {title}
  {{Sivers Effect for Pion and Kaon Production in Semi-Inclusive Deep Inelastic
  Scattering}},}\ }\href {\doibase 10.1140/epja/i2008-10697-y} {\bibfield
  {journal} {\bibinfo  {journal} {Eur. Phys. J. A}\ }\textbf {\bibinfo {volume}
  {39}},\ \bibinfo {pages} {89} (\bibinfo {year} {2009})}\BibitemShut {NoStop}%
\bibitem [{\citenamefont {Collins}\ \emph {et~al.}(2006)\citenamefont
  {Collins}, \citenamefont {Efremov}, \citenamefont {Goeke}, \citenamefont
  {Menzel}, \citenamefont {Metz},\ and\ \citenamefont
  {Schweitzer}}]{Collins:2005ie}%
  \BibitemOpen
  \bibfield  {author} {\bibinfo {author} {\bibfnamefont {J.~C.}\ \bibnamefont
  {Collins}}, \bibinfo {author} {\bibfnamefont {A.~V.}\ \bibnamefont
  {Efremov}}, \bibinfo {author} {\bibfnamefont {K.}~\bibnamefont {Goeke}},
  \bibinfo {author} {\bibfnamefont {S.}~\bibnamefont {Menzel}}, \bibinfo
  {author} {\bibfnamefont {A.}~\bibnamefont {Metz}}, \ and\ \bibinfo {author}
  {\bibfnamefont {P.}~\bibnamefont {Schweitzer}},\ }\bibfield  {title}
  {\enquote {\bibinfo {title} {{Sivers effect in semi-inclusive deeply
  inelastic scattering}},}\ }\href {\doibase 10.1103/PhysRevD.73.014021}
  {\bibfield  {journal} {\bibinfo  {journal} {Phys. Rev. D}\ }\textbf {\bibinfo
  {volume} {73}},\ \bibinfo {pages} {014021} (\bibinfo {year}
  {2006})}\BibitemShut {NoStop}%
\bibitem [{\citenamefont {Vogelsang}\ and\ \citenamefont
  {Yuan}(2005)}]{Vogelsang:2005cs}%
  \BibitemOpen
  \bibfield  {author} {\bibinfo {author} {\bibfnamefont {W.}~\bibnamefont
  {Vogelsang}}\ and\ \bibinfo {author} {\bibfnamefont {F.}~\bibnamefont
  {Yuan}},\ }\bibfield  {title} {\enquote {\bibinfo {title} {{Single-transverse
  spin asymmetries: From DIS to hadronic collisions}},}\ }\href {\doibase
  10.1103/PhysRevD.72.054028} {\bibfield  {journal} {\bibinfo  {journal} {Phys.
  Rev. D}\ }\textbf {\bibinfo {volume} {72}},\ \bibinfo {pages} {054028}
  (\bibinfo {year} {2005})}\BibitemShut {NoStop}%
\bibitem [{\citenamefont {Anselmino}\ \emph {et~al.}(2005)\citenamefont
  {Anselmino}, \citenamefont {Boglione}, \citenamefont {D'Alesio},
  \citenamefont {Kotzinian}, \citenamefont {Murgia},\ and\ \citenamefont
  {Prokudin}}]{Anselmino:2005ea}%
  \BibitemOpen
  \bibfield  {author} {\bibinfo {author} {\bibfnamefont {M.}~\bibnamefont
  {Anselmino}}, \bibinfo {author} {\bibfnamefont {M.}~\bibnamefont {Boglione}},
  \bibinfo {author} {\bibfnamefont {U.}~\bibnamefont {D'Alesio}}, \bibinfo
  {author} {\bibfnamefont {A.}~\bibnamefont {Kotzinian}}, \bibinfo {author}
  {\bibfnamefont {F.}~\bibnamefont {Murgia}}, \ and\ \bibinfo {author}
  {\bibfnamefont {A.}~\bibnamefont {Prokudin}},\ }\bibfield  {title} {\enquote
  {\bibinfo {title} {{Extracting the Sivers function from polarized SIDIS data
  and making predictions}},}\ }\href {\doibase 10.1103/PhysRevD.72.094007,
  10.1103/PhysRevD.72.099903} {\bibfield  {journal} {\bibinfo  {journal} {Phys.
  Rev. D}\ }\textbf {\bibinfo {volume} {72}},\ \bibinfo {pages} {094007}
  (\bibinfo {year} {2005})},\ \bibinfo {note} {[Erratum: Phys. Rev. D {\bf 72},
  099903(E) (2005)]}\BibitemShut {NoStop}%
\bibitem [{\citenamefont {Adare}\ \emph {et~al.}(2010)\citenamefont {Adare}
  \emph {et~al.}}]{Adare:2010bd}%
  \BibitemOpen
  \bibfield  {author} {\bibinfo {author} {\bibfnamefont {A.}~\bibnamefont
  {Adare}} \emph {et~al.} (\bibinfo {collaboration} {PHENIX Collaboration}),\
  }\bibfield  {title} {\enquote {\bibinfo {title} {{Measurement of Transverse
  Single-Spin Asymmetries for $J/\psi$ Production in Polarized $p+p$ Collisions
  at $\sqrt{s} = 200$ GeV}},}\ }\href {\doibase 10.1103/PhysRevD.82.112008,
  10.1103/PhysRevD.86.099904} {\bibfield  {journal} {\bibinfo  {journal} {Phys.
  Rev. D}\ }\textbf {\bibinfo {volume} {82}},\ \bibinfo {pages} {112008}
  (\bibinfo {year} {2010})},\ \bibinfo {note} {[Erratum: Phys. Rev. D {\bf 86},
  099904(E) (2012)]}\BibitemShut {NoStop}%
\bibitem [{\citenamefont {Godbole}\ \emph
  {et~al.}(2017{\natexlab{b}})\citenamefont {Godbole}, \citenamefont {Kaushik},
  \citenamefont {Misra}, \citenamefont {Rawoot},\ and\ \citenamefont
  {Sonawane}}]{Godbole:2017syo}%
  \BibitemOpen
  \bibfield  {author} {\bibinfo {author} {\bibfnamefont {R.~M.}\ \bibnamefont
  {Godbole}}, \bibinfo {author} {\bibfnamefont {A.}~\bibnamefont {Kaushik}},
  \bibinfo {author} {\bibfnamefont {A.}~\bibnamefont {Misra}}, \bibinfo
  {author} {\bibfnamefont {V.}~\bibnamefont {Rawoot}}, \ and\ \bibinfo {author}
  {\bibfnamefont {B.}~\bibnamefont {Sonawane}},\ }\bibfield  {title} {\enquote
  {\bibinfo {title} {{Transverse single spin asymmetry in $p+p^\uparrow
  \rightarrow J/\psi +X$}},}\ }\href {\doibase 10.1103/PhysRevD.96.096025}
  {\bibfield  {journal} {\bibinfo  {journal} {Phys. Rev. D}\ }\textbf {\bibinfo
  {volume} {96}},\ \bibinfo {pages} {096025} (\bibinfo {year}
  {2017}{\natexlab{b}})}\BibitemShut {NoStop}%
\bibitem [{\citenamefont {D'Alesio}\ \emph {et~al.}(2017)\citenamefont
  {D'Alesio}, \citenamefont {Murgia}, \citenamefont {Pisano},\ and\
  \citenamefont {Taels}}]{DAlesio:2017rzj}%
  \BibitemOpen
  \bibfield  {author} {\bibinfo {author} {\bibfnamefont {U.}~\bibnamefont
  {D'Alesio}}, \bibinfo {author} {\bibfnamefont {F.}~\bibnamefont {Murgia}},
  \bibinfo {author} {\bibfnamefont {C.}~\bibnamefont {Pisano}}, \ and\ \bibinfo
  {author} {\bibfnamefont {P.}~\bibnamefont {Taels}},\ }\bibfield  {title}
  {\enquote {\bibinfo {title} {{Probing the gluon Sivers function in
  $p^\uparrow p\to J/\psi\,X$ and $p^\uparrow p \to D\,X$}},}\ }\href {\doibase
  10.1103/PhysRevD.96.036011} {\bibfield  {journal} {\bibinfo  {journal} {Phys.
  Rev. D}\ }\textbf {\bibinfo {volume} {96}},\ \bibinfo {pages} {036011}
  (\bibinfo {year} {2017})}\BibitemShut {NoStop}%
\bibitem [{\citenamefont {Gelis}\ \emph {et~al.}(2010)\citenamefont {Gelis},
  \citenamefont {Iancu}, \citenamefont {Jalilian-Marian},\ and\ \citenamefont
  {Venugopalan}}]{Gelis:2010nm}%
  \BibitemOpen
  \bibfield  {author} {\bibinfo {author} {\bibfnamefont {F.}~\bibnamefont
  {Gelis}}, \bibinfo {author} {\bibfnamefont {E.}~\bibnamefont {Iancu}},
  \bibinfo {author} {\bibfnamefont {J.}~\bibnamefont {Jalilian-Marian}}, \ and\
  \bibinfo {author} {\bibfnamefont {R.}~\bibnamefont {Venugopalan}},\
  }\bibfield  {title} {\enquote {\bibinfo {title} {{The Color Glass
  Condensate}},}\ }\href {\doibase 10.1146/annurev.nucl.010909.083629}
  {\bibfield  {journal} {\bibinfo  {journal} {Ann. Rev. Nucl. Part. Sci.}\
  }\textbf {\bibinfo {volume} {60}},\ \bibinfo {pages} {463} (\bibinfo {year}
  {2010})}\BibitemShut {NoStop}%
\bibitem [{\citenamefont {McLerran}\ and\ \citenamefont
  {Venugopalan}(1994)}]{McLerran:1993ni}%
  \BibitemOpen
  \bibfield  {author} {\bibinfo {author} {\bibfnamefont {L.~D.}\ \bibnamefont
  {McLerran}}\ and\ \bibinfo {author} {\bibfnamefont {R.}~\bibnamefont
  {Venugopalan}},\ }\bibfield  {title} {\enquote {\bibinfo {title} {{Computing
  quark and gluon distribution functions for very large nuclei}},}\ }\href
  {\doibase 10.1103/PhysRevD.49.2233} {\bibfield  {journal} {\bibinfo
  {journal} {Phys. Rev. D}\ }\textbf {\bibinfo {volume} {49}},\ \bibinfo
  {pages} {2233--2241} (\bibinfo {year} {1994})}\BibitemShut {NoStop}%
\bibitem [{\citenamefont {Qiu}\ and\ \citenamefont {Vitev}(2006)}]{Qiu:2004da}%
  \BibitemOpen
  \bibfield  {author} {\bibinfo {author} {\bibfnamefont {J.}~\bibnamefont
  {Qiu}}\ and\ \bibinfo {author} {\bibfnamefont {I.}~\bibnamefont {Vitev}},\
  }\bibfield  {title} {\enquote {\bibinfo {title} {{Coherent QCD multiple
  scattering in proton-nucleus collisions}},}\ }\href {\doibase
  10.1016/j.physletb.2005.10.073} {\bibfield  {journal} {\bibinfo  {journal}
  {Phys. Lett. B}\ }\textbf {\bibinfo {volume} {632}},\ \bibinfo {pages} {507}
  (\bibinfo {year} {2006})}\BibitemShut {NoStop}%
\bibitem [{\citenamefont {Sch{\"a}fer}\ and\ \citenamefont
  {Zhou}(2014{\natexlab{a}})}]{Schafer:2014zea}%
  \BibitemOpen
  \bibfield  {author} {\bibinfo {author} {\bibfnamefont {A.}~\bibnamefont
  {Sch{\"a}fer}}\ and\ \bibinfo {author} {\bibfnamefont {J.}~\bibnamefont
  {Zhou}},\ }\bibfield  {title} {\enquote {\bibinfo {title} {{Transverse single
  spin asymmetry in direct photon production in polarized pA collisions}},}\
  }\href {\doibase 10.1103/PhysRevD.90.034016} {\bibfield  {journal} {\bibinfo
  {journal} {Phys. Rev. D}\ }\textbf {\bibinfo {volume} {90}},\ \bibinfo
  {pages} {034016} (\bibinfo {year} {2014}{\natexlab{a}})}\BibitemShut
  {NoStop}%
\bibitem [{\citenamefont {Sch{\"a}fer}\ and\ \citenamefont
  {Zhou}(2014{\natexlab{b}})}]{Schafer:2014xpa}%
  \BibitemOpen
  \bibfield  {author} {\bibinfo {author} {\bibfnamefont {A.}~\bibnamefont
  {Sch{\"a}fer}}\ and\ \bibinfo {author} {\bibfnamefont {J.}~\bibnamefont
  {Zhou}},\ }\bibfield  {title} {\enquote {\bibinfo {title} {{Color
  entanglement for $\gamma$-jet production in polarized $p$A collisions}},}\
  }\href {\doibase 10.1103/PhysRevD.90.094012} {\bibfield  {journal} {\bibinfo
  {journal} {Phys. Rev. D}\ }\textbf {\bibinfo {volume} {90}},\ \bibinfo
  {pages} {094012} (\bibinfo {year} {2014}{\natexlab{b}})}\BibitemShut
  {NoStop}%
\bibitem [{\citenamefont {Zhou}(2017{\natexlab{a}})}]{Zhou:2017sdx}%
  \BibitemOpen
  \bibfield  {author} {\bibinfo {author} {\bibfnamefont {J.}~\bibnamefont
  {Zhou}},\ }\bibfield  {title} {\enquote {\bibinfo {title} {{Single spin
  asymmetries in forward p-p/A collisions revisited: the role of color
  entanglement}},}\ }\href {\doibase 10.1103/PhysRevD.96.034027} {\bibfield
  {journal} {\bibinfo  {journal} {Phys. Rev. D}\ }\textbf {\bibinfo {volume}
  {96}},\ \bibinfo {pages} {034027} (\bibinfo {year}
  {2017}{\natexlab{a}})}\BibitemShut {NoStop}%
\bibitem [{\citenamefont {Zhou}(2017{\natexlab{b}})}]{Zhou:2017mpw}%
  \BibitemOpen
  \bibfield  {author} {\bibinfo {author} {\bibfnamefont {J.}~\bibnamefont
  {Zhou}},\ }\bibfield  {title} {\enquote {\bibinfo {title} {{Color
  entanglement like effect in collinear twist-3 factorization}},}\ }\href
  {\doibase 10.1103/PhysRevD.96.114001} {\bibfield  {journal} {\bibinfo
  {journal} {Phys. Rev. D}\ }\textbf {\bibinfo {volume} {96}},\ \bibinfo
  {pages} {114001} (\bibinfo {year} {2017}{\natexlab{b}})}\BibitemShut
  {NoStop}%
\bibitem [{\citenamefont {Akikawa}\ \emph {et~al.}(2003)\citenamefont {Akikawa}
  \emph {et~al.}}]{PHENIX:2003}%
  \BibitemOpen
  \bibfield  {author} {\bibinfo {author} {\bibfnamefont {H.}~\bibnamefont
  {Akikawa}} \emph {et~al.} (\bibinfo {collaboration} {PHENIX Collaboration}),\
  }\bibfield  {title} {\enquote {\bibinfo {title} {{PHENIX muon arms}},}\
  }\href {\doibase 10.1016/S0168-9002(02)01955-1} {\bibfield  {journal}
  {\bibinfo  {journal} {Nucl. Instrum. Methods Phys. Res., Sec. A}\ }\textbf
  {\bibinfo {volume} {499}},\ \bibinfo {pages} {537--548} (\bibinfo {year}
  {2003})}\BibitemShut {NoStop}%
\bibitem [{\citenamefont {Adcox}\ \emph
  {et~al.}(2003{\natexlab{a}})\citenamefont {Adcox} \emph
  {et~al.}}]{Adcox:2003zm}%
  \BibitemOpen
  \bibfield  {author} {\bibinfo {author} {\bibfnamefont {K.}~\bibnamefont
  {Adcox}} \emph {et~al.} (\bibinfo {collaboration} {PHENIX Collaboration}),\
  }\bibfield  {title} {\enquote {\bibinfo {title} {{PHENIX detector
  overview}},}\ }\href {\doibase 10.1016/S0168-9002(02)01950-2} {\bibfield
  {journal} {\bibinfo  {journal} {Nucl. Instrum. Meth. A}\ }\textbf {\bibinfo
  {volume} {499}},\ \bibinfo {pages} {469--479} (\bibinfo {year}
  {2003}{\natexlab{a}})}\BibitemShut {NoStop}%
\bibitem [{\citenamefont {Adachi}\ \emph {et~al.}(2013)\citenamefont {Adachi}
  \emph {et~al.}}]{Adachi:2013qha}%
  \BibitemOpen
  \bibfield  {author} {\bibinfo {author} {\bibfnamefont {S.}~\bibnamefont
  {Adachi}} \emph {et~al.},\ }\bibfield  {title} {\enquote {\bibinfo {title}
  {{Trigger electronics upgrade of PHENIX muon tracker}},}\ }\href {\doibase
  10.1016/j.nima.2012.11.088} {\bibfield  {journal} {\bibinfo  {journal} {Nucl.
  Instrum. Meth. A}\ }\textbf {\bibinfo {volume} {703}},\ \bibinfo {pages}
  {114--132} (\bibinfo {year} {2013})}\BibitemShut {NoStop}%
\bibitem [{\citenamefont {Adcox}\ \emph
  {et~al.}(2003{\natexlab{b}})\citenamefont {Adcox} \emph
  {et~al.}}]{PHENIX_FULL:2003}%
  \BibitemOpen
  \bibfield  {author} {\bibinfo {author} {\bibfnamefont {K.}~\bibnamefont
  {Adcox}} \emph {et~al.} (\bibinfo {collaboration} {PHENIX Collaboration}),\
  }\bibfield  {title} {\enquote {\bibinfo {title} {{PHENIX detector
  overview}},}\ }\href {\doibase 10.1016/S0168-9002(02)01950-2} {\bibfield
  {journal} {\bibinfo  {journal} {Nucl. Instrum. Methods Phys. Res., Sec. A}\
  }\textbf {\bibinfo {volume} {499}},\ \bibinfo {pages} {469--479} (\bibinfo
  {year} {2003}{\natexlab{b}})}\BibitemShut {NoStop}%
\bibitem [{\citenamefont {Adler}\ \emph {et~al.}(2003)\citenamefont {Adler}
  \emph {et~al.}}]{Adler:2003pb}%
  \BibitemOpen
  \bibfield  {author} {\bibinfo {author} {\bibfnamefont {S.~S.}\ \bibnamefont
  {Adler}} \emph {et~al.} (\bibinfo {collaboration} {PHENIX Collaboration}),\
  }\bibfield  {title} {\enquote {\bibinfo {title} {{Mid-rapidity neutral pion
  production in proton proton collisions at $\sqrt{s}$ = 200-GeV}},}\ }\href
  {\doibase 10.1103/PhysRevLett.91.241803} {\bibfield  {journal} {\bibinfo
  {journal} {Phys. Rev. Lett.}\ }\textbf {\bibinfo {volume} {91}},\ \bibinfo
  {pages} {241803} (\bibinfo {year} {2003})}\BibitemShut {NoStop}%
\bibitem [{\citenamefont {Adare}\ \emph
  {et~al.}(2014{\natexlab{c}})\citenamefont {Adare} \emph
  {et~al.}}]{Adare:2013nff}%
  \BibitemOpen
  \bibfield  {author} {\bibinfo {author} {\bibfnamefont {A.}~\bibnamefont
  {Adare}} \emph {et~al.} (\bibinfo {collaboration} {PHENIX Collaboration}),\
  }\bibfield  {title} {\enquote {\bibinfo {title} {{Centrality categorization
  for $R_{p(d)+A}$ in high-energy collisions}},}\ }\href {\doibase
  10.1103/PhysRevC.90.034902} {\bibfield  {journal} {\bibinfo  {journal} {Phys.
  Rev. C}\ }\textbf {\bibinfo {volume} {90}},\ \bibinfo {pages} {034902}
  (\bibinfo {year} {2014}{\natexlab{c}})}\BibitemShut {NoStop}%
\bibitem [{\citenamefont {Alekseev}(2006)}]{Alekseev:1489159}%
  \BibitemOpen
  \bibfield  {author} {\bibinfo {author} {\bibfnamefont {I}~\bibnamefont
  {Alekseev}},\ }\href {http://inspirehep.net/record/1614081} {\emph {\bibinfo
  {title} {{Configuration Manual: Polarized Proton Collider at RHIC}}}},\
  \bibinfo {type} {Tech. Rep.}\ (\bibinfo  {institution} {Brookhaven National
  Laboratory},\ \bibinfo {address} {Upton, NY},\ \bibinfo {year}
  {2006})\BibitemShut {NoStop}%
\bibitem [{\citenamefont {Adare}\ \emph
  {et~al.}(2016{\natexlab{a}})\citenamefont {Adare} \emph
  {et~al.}}]{Adare:2015ozj}%
  \BibitemOpen
  \bibfield  {author} {\bibinfo {author} {\bibfnamefont {A.}~\bibnamefont
  {Adare}} \emph {et~al.} (\bibinfo {collaboration} {PHENIX Collaboration}),\
  }\bibfield  {title} {\enquote {\bibinfo {title} {{Inclusive cross section and
  double-helicity asymmetry for $\pi^{0}$ production at midrapidity in
  $p$$+$$p$ collisions at $\sqrt{s}=510$ GeV}},}\ }\href {\doibase
  10.1103/PhysRevD.93.011501} {\bibfield  {journal} {\bibinfo  {journal} {Phys.
  Rev. D}\ }\textbf {\bibinfo {volume} {93}},\ \bibinfo {pages} {011501}
  (\bibinfo {year} {2016}{\natexlab{a}})}\BibitemShut {NoStop}%
\bibitem [{\citenamefont {Adare}\ \emph
  {et~al.}(2016{\natexlab{b}})\citenamefont {Adare} \emph
  {et~al.}}]{Adare:2016cqe}%
  \BibitemOpen
  \bibfield  {author} {\bibinfo {author} {\bibfnamefont {A.}~\bibnamefont
  {Adare}} \emph {et~al.} (\bibinfo {collaboration} {PHENIX Collaboration}),\
  }\bibfield  {title} {\enquote {\bibinfo {title} {{Measurements of
  double-helicity asymmetries in inclusive $J/\psi$ production in
  longitudinally polarized $p+p$ collisions at $\sqrt{s}=510$ GeV}},}\ }\href
  {\doibase 10.1103/PhysRevD.94.112008} {\bibfield  {journal} {\bibinfo
  {journal} {Phys. Rev. D}\ }\textbf {\bibinfo {volume} {94}},\ \bibinfo
  {pages} {112008} (\bibinfo {year} {2016}{\natexlab{b}})}\BibitemShut
  {NoStop}%
\bibitem [{\citenamefont {MacKay}(2003)}]{MacKay}%
  \BibitemOpen
  \bibfield  {author} {\bibinfo {author} {\bibfnamefont {D.~J.}\ \bibnamefont
  {MacKay}},\ }\href@noop {} {\emph {\bibinfo {title} {Information Theory,
  Inference and Learning Algorithms}}}\ (\bibinfo  {publisher} {Cambridge
  University Press},\ \bibinfo {year} {2003})\BibitemShut {NoStop}%
\bibitem [{\citenamefont {Rasmussen}\ and\ \citenamefont
  {Williams}(2006)}]{Rasmussen}%
  \BibitemOpen
  \bibfield  {author} {\bibinfo {author} {\bibfnamefont {C.~E.}\ \bibnamefont
  {Rasmussen}}\ and\ \bibinfo {author} {\bibfnamefont {C.~K.~I.}\ \bibnamefont
  {Williams}},\ }\href@noop {} {\emph {\bibinfo {title} {Gaussian Processes for
  Machine Learning}}},\ edited by\ \bibinfo {editor} {\bibfnamefont
  {T.}~\bibnamefont {Dietterich}}\ (\bibinfo  {publisher} {MIT Press},\
  \bibinfo {year} {2006})\BibitemShut {NoStop}%
\bibitem [{\citenamefont {Lauritzen}(1996)}]{Lauritzen}%
  \BibitemOpen
  \bibfield  {author} {\bibinfo {author} {\bibfnamefont {S.~L.}\ \bibnamefont
  {Lauritzen}},\ }\href@noop {} {\emph {\bibinfo {title} {Graphical models}}}\
  (\bibinfo  {publisher} {Clarendon Press Oxford University Press},\ \bibinfo
  {year} {1996})\BibitemShut {NoStop}%
\bibitem [{\citenamefont {Barber}(2012)}]{barber2012bayesian}%
  \BibitemOpen
  \bibfield  {author} {\bibinfo {author} {\bibfnamefont {D.}~\bibnamefont
  {Barber}},\ }\href@noop {} {\emph {\bibinfo {title} {Bayesian reasoning and
  machine learning}}}\ (\bibinfo  {publisher} {Cambridge University Press},\
  \bibinfo {year} {2012})\BibitemShut {NoStop}%
\bibitem [{\citenamefont {Adare}\ \emph
  {et~al.}(2016{\natexlab{c}})\citenamefont {Adare} \emph
  {et~al.}}]{Adare:2015gsd}%
  \BibitemOpen
  \bibfield  {author} {\bibinfo {author} {\bibfnamefont {A.}~\bibnamefont
  {Adare}} \emph {et~al.} (\bibinfo {collaboration} {PHENIX Collaboration}),\
  }\bibfield  {title} {\enquote {\bibinfo {title} {{Measurement of
  parity-violating spin asymmetries in $W^{\pm}$ production at midrapidity in
  longitudinally polarized $p$$+$$p$ collisions}},}\ }\href {\doibase
  10.1103/PhysRevD.93.051103} {\bibfield  {journal} {\bibinfo  {journal} {Phys.
  Rev. D}\ }\textbf {\bibinfo {volume} {93}},\ \bibinfo {pages} {051103}
  (\bibinfo {year} {2016}{\natexlab{c}})}\BibitemShut {NoStop}%
\bibitem [{\citenamefont {Skwarnicki}(1986)}]{Skwarnicki:1986xj}%
  \BibitemOpen
  \bibfield  {author} {\bibinfo {author} {\bibfnamefont {T.}~\bibnamefont
  {Skwarnicki}},\ }\emph {\bibinfo {title} {{A study of the radiative CASCADE
  transitions between the Upsilon-Prime and Upsilon resonances}}},\ \href
  {http://lss.fnal.gov/cgi-bin/find_paper.pl?other/thesis/skwarnicki.pdf}
  {Ph.D. thesis},\ \bibinfo  {school} {Cracow, INP} (\bibinfo {year}
  {1986})\BibitemShut {NoStop}%
\bibitem [{\citenamefont {Aidala}\ \emph {et~al.}(2018)\citenamefont {Aidala}
  \emph {et~al.}}]{Aidala:2017cnz}%
  \BibitemOpen
  \bibfield  {author} {\bibinfo {author} {\bibfnamefont {C.}~\bibnamefont
  {Aidala}} \emph {et~al.} (\bibinfo {collaboration} {PHENIX Collaboration}),\
  }\bibfield  {title} {\enquote {\bibinfo {title} {{Nuclear Dependence of the
  Transverse-Single-Spin Asymmetry for Forward Neutron Production in Polarized
  $p+A$ Collisions at $\sqrt{s_{NN}}=200$ GeV}},}\ }\href {\doibase
  10.1103/PhysRevLett.120.022001} {\bibfield  {journal} {\bibinfo  {journal}
  {Phys. Rev. Lett.}\ }\textbf {\bibinfo {volume} {120}},\ \bibinfo {pages}
  {022001} (\bibinfo {year} {2018})}\BibitemShut {NoStop}%
\end{thebibliography}

%
 
\end{document}